\documentclass[journal,10pt,twocolumn]{IEEEtran}
\RequirePackage[final]{graphicx}
\usepackage{amsmath}
\usepackage{caption}
\usepackage{subcaption}
\usepackage{cite}

\usepackage[table,xcdraw]{xcolor}
\usepackage{multirow}
\usepackage[table,xcdraw]{xcolor}

\usepackage{algorithm}
\usepackage{algorithmic}
\usepackage{amsfonts}
\usepackage{amsthm}

\usepackage{url}

\ifCLASSINFOpdf
\else
\fi

\begin{document}
\title{Joint Range and Angle Estimation for FMCW MIMO Radar and Its Application}


\author{Junghoon~Kim,
        Joohwan~Chun,
        and~Sungchan Song%
        % <-this % stops a space

%\author{Michael~Shell,~\IEEEmembership{Member,~IEEE,}
%        John~Doe,~\IEEEmembership{Fellow,~OSA,}
%        and~Jane~Doe,~\IEEEmembership{Life~Fellow,~IEEE}% <-this % stops a space
\thanks{Junghoon Kim and Joohwan Chun are with Department
of Electrical Engineering, Korea Advanced Institute of Science and Technology, Daejeon, South Korea
and Sungchan Song, with Hanwha Systems, Yongin City, South Korea.}
\thanks{This work was supported in part by Hanwha Systems under the industry-academia collaboration
program in 2017.}}
\maketitle


\begin{abstract}
Recently, frequency-modulated continuous wave (FMCW) radars with array antennas are gaining in popularity
on a wide variety of commercial applications.
A usual approach of the range and angle estimation of a target with an array FMCW radar is to
form a range-angle matrix with deramped receive signal, and then apply the
two-dimensional fast Fourier transformation (2D-FFT) on the range-angle matrix.
However, such frequency estimation approaches give bias error because
two frequencies on the range-angle matrix are not independent to each other,
unlike the 2D angle estimation with a passive planar antenna array.
We propose a new maximum-likelihood based algorithm for the joint range
and angle estimation of multiple targets with an array FMCW radar, and show that
the proposed algorithm achieves the
Cramer-Rao bounds (CRBs) both for the range and angle estimation.
The proposed algorithm is also compared with other algorithms for
a simultaneous localization and mapping (SLAM) problem.
%which leads to bias error.
%
%Also, in case of multiple targets, the range and angle estimation performances are degraded
%because of interferences from other targets.
%In this paper, we address those problems for accurate range and angle estimation.
%Furthermore, we also derive Cramer-Rao bounds (CRBs) of range and angle, and compare
%the proposed algorithm against conventional approaches as well as CRB.
\end{abstract}

\begin{IEEEkeywords}
FMCW radars, maximum likelihood estimation, 2D-FFT, 2D-MUSIC, CRB, joint range and angle estimation
\end{IEEEkeywords}

% \fi
\IEEEpeerreviewmaketitle

%\let\oldsection\subsection
%\renewcommand\subsection{\clearpage\oldsection}

%%%%%%%%%%%%%%%%%%%%%%%%%%%%%%%%%%%%%%%%%%%%%%%%%%%%%%%
% Section 1. Introduction
%%%%%%%%%%%%%%%%%%%%%%%%%%%%%%%%%%%%%%%%%%%%%%%%%%%%%%%
\section{Introduction}
\label{sec:intro}

%%%%% The use of FMCW radar
Frequency-modulated continuous wave (FMCW) radars are widely used in short-range applications such as altimeters
\cite{skolnik1962introduction},
automotive radars
\cite{Hasch2012,Rohling2001,Schneider2005}
% ,Winkler2007
and more recently,
synthetic aperture radars (SARs)
\cite{Meta2007,Wang2013,Wang2010,giusti2011range,liu2013bistatic}.
% DeWit2006
Advantages of FMCW radars lie in their light weight, low power consumption and cost-effectiveness,
while achieving relatively high range resolution \cite{Stove1993a,brennan2011determination}.
FMCW radars may have a full receive (Rx) antenna array or a multiple-input multiple-output (MIMO)
virtual antenna array~\cite{babur2013nearly,Feger2009a,Belfiori2012a,majumder2016design}.
Either with the full Rx array or with the MIMO virtual array,
we first form
a range-angle matrix,
where the angle dimension may extend to the size of the virtual array in case of the MIMO system.
Then, to get range and angle estimates of each target, we may apply
two-dimensional (2D) frequency estimation algorithms on the range-angle matrix,
such as the 2D fast Fourier transformation (2D-FFT) \cite{Feger2009a},
joint angle-frequency estimation (JAFE) \cite{Lemma2003},
2D multiple signal classification (2D-MUSIC) \cite{Belfiori2012a},
or the joint discrete Fourier transform (DFT)-estimation of signal parameters via rotational
invariance techniques (ESPRIT) \cite{Kim2015a}.

However, these traditional 2D frequency estimation algorithms
ignore the fact that
two frequencies on each domain are dependent on each other and are coupled.
This is because
the frequency on the range domain, which is proportional to the range to the target,
differs slightly at each antenna,
%with the length of $d \sin \theta$,
%where $d$ is distance between adjacent antennas and $\theta$ is an
depending on the angle to the target.
Most previous works do not include the range difference per antenna in the signal model,
resulting in inevitable bias error.
%occurs when the conventional 2D frequency estimation method is applied.
%
%%%%%% Conventional Approaches : LSE
%
We remark that the least squares estimation algorithm in \cite{Feger2008}
can handle the
%is introduced
%considering
range difference of antennas, and yields an accurate estimation in case of a single target.
%In this method, a grid search on range-angle domain is performed by constructing the cost function for each target, and it
However with multiple targets, interference from other targets degrades the performance.
%Also, its being based on the grid search, the computational complexity
%is prohibitively large
%to determine the pair of range and angle,
%when high estimation accuracy is needed.

%%%%% Proposed method
We propose a maximum-likelihood algorithm for the joint range and angle estimation of multiple targets
with FMCW radars, to solve the aforementioned bias and interference error problem.
We also derive the Cramer-Rao bounds (CRBs) of the estimates, and compare the proposed algorithm with
other algorithms as well as the CRB.
%
% % Applicable to general case
We remark that the proposed algorithm is applicable not only to the range-angle domain
but to the ordinary range-Doppler or to the angle-Doppler domain of the space-time adaptive processing (STAP)~\cite{srra}.
This is because we similarly have
%the range difference per antenna on the range-angle domain is identical
%to
the range difference along the pulse dimension on the range-Doppler domain, and the angle difference
along the pulse dimension on the angle-Doppler domain.
%
%For example,
%
%in FMCW SAR \cite{meta2007signal,DeWit2006,Wang2013} or FMCW MIMO \cite{Guetlein2013c}, range
%migration occurs in range-Doppler domain due to Doppler,
%and the range bin difference depends on the parameters;
%The range bin difference per pulse can be more than 1 bin for airborne FMCW SAR \cite{meta2007signal},
%or less than 0.1 bin for automobile FMCW SAR \cite{Wang2013}.
%
%Normally, the range migration is corrected by multiplying raw data by Doppler compensation function,
%and 2D frequency estimation is performed on the modified range-Doppler domain.
%
%This is suitable for single targets, but not for multiple targets, because the Doppler compensation
%function with one target's Doppler information is multiplied by the raw data containing all targets'
%responses, which leads to an interference error.
%Therefore, in the case of multiple targets, the proposed algorithm can be extended to range-Doppler domain
%to handle the interference between targets’ response.
%if the range bin difference is not that large.

The proposed algorithm would be of benefit to the radar-based simultaneous
localization and mapping (SLAM), which updates the map of unknown environment, while simultaneous keeping
track of own vehicle autonomously~\cite{amjv, lcjl}, for short-distance autonomous navigation.
%
%SLAM technique is useful when the global positioning system (GPS) is
%not accurate enough or unavailable, or also when environment map is not well-defined or
%unavailable.
%With radar-based SLAM, the vehicle
%measures the ranges and angles to nearby opportunistic objects, when the vehicle
%is about to move from the exactly known origin.
%
If the vehicle carries an on-board environmental map, it may use the
global positioning system (GPS) receiver or its own active sensor such
as radar \cite{ng} to localize itself.
However, the GPS signal is not accurate enough for precision navigation, and is not even available in indoor
environment.
Also the map is not very useful for non-static environment such as with moving objects.
To localize the own vehicle relative to the environment, the radar-based SLAM utilizes the range and angle measurements
to a set of scatterers, called a point cloud, of a nearby opportunistic objects, as explained briefly below.

First, consider a vehicle that is standing at the exactly known origin.
The radar in the vehicle measures the ranges and angles to the scatterers of objects nearby.
These measurements will fix the position of those
scatterers. When the vehicle moves to a new position after a very short period of time, relative
positions of those scatterers seen from the vehicle will alter.
Then by comparing the two sets of relative positions of the scatterers, before and after the movement,
the vehicle is able to determine its new location. This new location becomes the new origin, and by repeating the
above process the vehicle can keep track of its trajectory relative to the nearby objects.
%With the radar sensor a set of scatterers, called a {\it point cloud}, will constitute the aformentioned objects.
%
Notice that measurement error in the range and angle will accumulate during the SLAM procedure, and therefore
the final position error of the vehicle can become quite large even with small error in the range and angle estimates.

In Sec 2, we shall show that the traditional 2D frequency estimation algorithms will give
bias error in the estimated range and angle.
A new maximum likelihood estimation algorithm is presented in Sec 3, and its
performance and comparison with other algorithms are given in Sec 4.
In Sec 5,
we shall show that SLAM utilizing the iterative closest point (ICP) method
with the range and angle estimates provided by the proposed algorithm will be able
to localize the own vehicle accurately during the
autonomous valet parking procedure~\cite{Paromtchik1996, Paromtchik, Hsu2008}, which is a special case of SLAM.
Conclusions are made in Sec 6.

\section{Estimation error of conventional approaches}
\subsection{FMCW radar signal model}

\label{sec:signalmodel}

The received deramped FMCW radar signal at a single antenna, for a single target, at time $n$ is given by
\begin{gather}
    s[n] = a{e^{j\phi }}{e^{j(2\pi \gamma \tau {T_s}n + 2\pi {f_c}\tau  - \pi \gamma {\tau ^2})}}, \quad 0 \le n \le N-1,
\end{gather}
where
$ae^{j\phi}$ is the reflectivity of the target,
$f_c$ is the operating frequency,
$\gamma$ is the chirp rate,
$T_s$ is the sampling period, %of analog digital converter (ADC),
$\tau$ is the time delay from the target to the receive antenna,
and $N$ is the total number of samples.

Let us assume that there are $M=PQ$ virtual receive antennas, where $P$ and $Q$ are the numbers of transmit and receive antennas,
respectively, and that
the array is uniform and linear with the inter-antenna spacing of
$d= \lambda/2$, where $\lambda = c/f_c$ is the wavelength of the signal.
Further, assuming that
there are $K$ targets, let us denote the incident angle of the signal from the $k$th target, $1 \le k \le K$
by $\theta_k$,
and the distance from the origin to the target, by $r_k$.
%Then the time delay to the $m$th antenna, $0 \le m \le M-1$ from the $k$th target will be $\tau_k[m] = (2r_k + mu_k)/c$,
%where $c$ is the speed of light and $u_k = d/sin \theta_k$
%is the signal path difference between two adjacent antennas.
Then the time delay $\tau_k[m]$
to the $m$th antenna, $0 \le m \le M-1$ from the $k$th target will be
\begin{gather}
\tau_k[m] = (2r_k + mu_k)/c, \quad u_k = d \sin\theta_k,
\end{gather}
where $c$ is the speed of light and
%%begin{gather}
%u_k = d sin\theta_k
%\end{gather}
$u_k$ is the signal path-length difference between two adjacent antennas.

Therefore, the deramped signal at the $m$th antenna,
reflected from the $k$th target will be
\begin{align}
{s_k}[n,m] &=
{a_k}{e^{j({\phi _k} - \pi \gamma {\tau _k^2}{{[m]}})}}{e^{j(2\pi {f_c}{\tau _k}[m] +
2\pi \gamma {\tau _k}[m]{T_s}n))}}
%\nonumber
\\
&= {a_k}{e^{j({\phi _k} - \pi \gamma {\tau _k^2}{{[m]}} + 4\pi {f_c}{r_k}/c)}}
\nonumber
\\
&\hspace{0.5cm} \times
{e^{j2\pi {f_c}m{u_k}/c}}{e^{j2\pi \gamma (2{r_k} + m{u_k})/c{T_s}n}}
%\nonumber
\\
&\approx {a_k}{e^{j{\psi _k}}}{e^{j2\pi \frac{{{u_k}}}{\lambda }m}}{e^{j2\pi (2{r_k} + m{u_k})\frac{B}{{cN}}n}},
\label{s_nm}
\end{align}
where in (\ref{s_nm}) we have used the relationship between the chirp rate $\gamma$ and the bandwidth
of the signal $B$; $\gamma = B/(NT_s)$
and the approximation that $\pi \gamma \tau_k^2 [m]$ is not dependent on $m$,
resulting in
%we approximate
\begin{gather}
\psi_k \approx {\phi _k} - \pi \gamma {\tau_k^2}{{[m]}} + 4\pi {f_c}{r_k}/c.
\end{gather}
Now the measurement at time $n$ and at antenna $m$ is the noisy sum of the signals from all targets,
\begin{gather}
    z[n,m] = \sum\limits_{k = 1}^K {{s_k}[n,m]}  + w[n,m],
    \label{eq_m}
\end{gather}
%
%where $K$ is the number of targets,
%${s_k}[n,m]$ is the demodulated FMCW signal of $k$-th target,
where $w[n,m] \sim CN(0,{\sigma ^2}) $ is complex additive Gaussian noise.

% Section 3-1.
\subsection{Bias error in the two-dimensional frequency estimation algorithms}

Two-dimensional frequency estimation algorithms, such as
2D-FFT \cite{Feger2009a} and
%JAFE \cite{Lemma2003},
2D-MUSIC \cite{Belfiori2012a}
%or joint DFT-ESPRIT \cite{Kim2015},
may be applied to get the range and angle estimates from the measurement (\ref{eq_m}). %in FMCW radars system.
%In order to improve the apparent resolution, zero-paddings are added.
%Nevertheless, bias error of range and angle occurs.
We shall show that such methods always give bias error for both range and angle estimates,
even with noiseless measurement and under infinite number of zero-padding, i.e., under the discrete-time Fourier
transformation (DTFT).

Let us consider the measurement (\ref{eq_m}) under noiseless condition
for a single target, with the definition of
${y_\theta } = Mu/\lambda = (M\sin\theta)/2$,
${x_r} = 2Br/c$,
and
${x_\theta } = Bu/c = (B\sin\theta)/(2f_c)$.
Then
\begin{align}
z[n,m] &= a{e^{j\psi }}{e^{j2\pi \frac{u}{\lambda }m}}{e^{j2\pi (2r + mu)\frac{B}{{cN}}n}} \\
&= a{e^{j\psi }}{e^{j2\pi \frac{{{y_\theta }}}{M}m}}{e^{j2\pi \frac{{{x_r} + m {x_\theta }}}{N}n}}.
\label{eq_x_theta}
\end{align}
In the below, it will be shown that the coupling term $x_\theta$, presenting in the measurement is responsible for the bias
of the estimate.
Let us consider the DTFT,
\begin{align}
S(x,y) = \sum\limits_{n = 0}^{N - 1} {\sum\limits_{m = 0}^{M - 1} {z[n,m]{e^{ - j2\pi \frac{x}{N}n}}
{e^{ - j2\pi \frac{y}{M}m}}} }.
\label{Sxy}
\end{align}

Our goal is to determine
$(x',y') = \arg \max |S(x,y)|$, and then, using
$x^\prime = 2Br^\prime/c$ and $y^\prime = M \sin\theta^\prime/2$,
compute $r^\prime$ and $\theta^\prime$,
which are to be compared with the
true $r$ and $\theta$.
Notice that
{\small
{
\begin{align}
|S(x',y')| &= a|\sum\limits_{n = 0}^{N - 1} {\sum\limits_{m = 0}^{M - 1} {{e^{j2\pi \frac{{({x_r} +
m {x_\theta } - x')}}{N}n}}{e^{j2\pi \frac{{{y_\theta } - y'}}{M}m}}} } |
%\nonumber
\label{S_1}
\\
&= a|\sum\limits_{m = 0}^{M - 1} {}\frac{{\sin (\pi ({x_r} + mx_\theta - x'))}}{{\sin (\pi ({x_r} + mx_\theta - x')/N)}}
%\nonumber
\\
&\hspace{1.5cm} \times
{e^{j\pi \frac{{N - 1}}{N}({x_r} + mx_\theta - x')}}{e^{j2\pi \frac{{{y_\theta } - y'}}{M}m}} |
\nonumber
\\
&= a|\sum\limits_{m = 0}^{M - 1} {}\frac{{\sin (\pi ({x_r} + mx_\theta - x'))}}{{\sin (\pi ({x_r} + mx_\theta - x')/N)}}
%\nonumber
%\\
%& \hspace{2cm} \times
{e^{j\pi \frac{{N - 1}}{N}mx_\theta}}{e^{j2\pi \frac{{{y_\theta } - y'}}{M}m}} |.
%\nonumber
\label{S_sin}
\end{align}
}}
The first factor in (\ref{S_sin}) can be simplified further by using the following inequalities,
which are proven in the sequel,
\begin{align}
\frac{\pi }{N}\mathop {\max }\limits_m |{x_r} + mx_\theta - x'|
& \le \frac{\pi }{N} (M - 1)|x_\theta|
\label{inequality_1}
\\
& \le \frac{\pi }{N} (M - 1)\frac{B}{{2{f_c}}}
\label{inequality_2}
\\
& \ll 1.
\label{inequality_3}
\end{align}
We know $x^\prime$ must be close to the true value $x_r$, and therefore
$x^\prime \in [{x_r},{x_r} + (M - 1)x_\theta]$ when $x_\theta \ge 0$,
or $x^\prime \in [{x_r} + (M - 1)x_\theta,{x_r}]$ when $x_\theta < 0$,
which assures the inequality (\ref{inequality_1}).
In addition, $N \gg M$ and $f_c \gg B$ hold commonly, which makes the inequality (\ref{inequality_3}) valid.
Therefore, using $(N-1)/N \approx 1$ in addition, (\ref{S_sin}) can be approximated as
{\small{
\begin{align}
|S(x^\prime, y^\prime)| &\approx aN|\sum\limits_{m = 0}^{M - 1} {\frac{{\sin (\pi ({x_r} + mx_\theta - x'))}}
{{\pi ({x_r} + mx_\theta - x')}}{e^{j\pi mx_\theta}}{e^{j2\pi \frac{{{y_\theta } - y'}}{M}m}}} |
%\nonumber
%\label{S_reduced_sin}
\\
&= aN|\sum\limits_{m = 0}^{M - 1} {{\rm{sinc(}}{x_r} + mx_\theta - x'{\rm{)}}{e^{j2\pi \frac{{{y_\theta } +
Mx_\theta/2 - y'}}{M}m}}} |.
%\nonumber
\label{S_sinc}
\end{align}
}}%
Furthermore, if $(M-1)B/(2f_c) < 1$, which is mostly the case, then we have
$\mathop {\max }\limits_m |{x_r} + mx_\theta - x'| < 1$ from (\ref{inequality_2}), which leads
${\rm{sinc(}}{x_r} + mx_\theta - x'{\rm{) > 0}}$.
Therefore (\ref{S_sinc}) becomes the maximum when the exponent of the complex exponential function is zero, i.e.,
\begin{gather}
    y' = {y_\theta } + \frac{M}{2}x_\theta.
    \label{eq_y'}
\end{gather}
Now by inserting
$y'=(M\sin\theta')/2$,
${y_\theta } = (M\sin\theta)/2 $, and
$x_\theta = (B\sin\theta)/(2f_c)$
into (\ref{eq_y'}),
we get
\begin{gather}
    \theta ' = \sin{^{ - 1}}((1 + \frac{B}{{2{f_c}}})\sin \theta ).
    \label{eq_theta'}
\end{gather}
With this $y^\prime$ in (\ref{eq_y'}), (\ref{S_sinc}) reduces to
\begin{gather}
    |S(x',y')| = aN|\sum\limits_{m = 0}^{M - 1} {{\rm{sinc(}}{x_r} + mx_\theta - x'{\rm{)}}} |.
    \label{S_x}
\end{gather}
Noting that
${{\rm{sinc(}}{x_r} + mx_\theta - x'{\rm{)}}} > 0$, (\ref{S_x}) will be the maximum when all $M$ points of
${{\rm{sinc(}}{x_r} + mx_\theta - x'{\rm{)}}}$ for $0 \le m \le M-1$, are symmetrically placed around the peak.
Therefore
\begin{gather}
    x' = {x_r} + \frac{{M - 1}}{2}x_\theta.
\label{eq_x'}
\end{gather}
By inserting
$x'=2Br'/c$, ${x_r} = 2Br/c$, and $x_\theta = (B\sin\theta)/(2f_c)$ into (\ref{eq_x'}), we get
\begin{gather}
    r' = r + \frac{{(M - 1)}}{8} \lambda \sin \theta.
    \label{eq_r'}
\end{gather}
Finally, from (\ref{eq_theta'}) and (\ref{eq_r'}), the bias errors of the angle and range
estimates are given by
\begin{align}
    {r^{(b)}} &= r' - r = \frac{{(M - 1)\lambda }}{8}\sin \theta,
    \label{eq_r'b}
    \\
    {\theta ^{(b)}} &= \theta ' - \theta  = \sin^{ - 1}((1 + \frac{B}{{2{f_c}}})\sin \theta ) - \theta.
    \label{eq_theta'b}
\end{align}
%
%%%%%%%%%%%%%%%%%%%%%%
%%%%%%%%%%%%%%%%%%%%%%
\subsection{The least square estimation algorithm}
\label{sec:LSE}
%Actually, 2D frequency estimation is not proper solution for accurate estimation of range and angle
%because of the term, $x_\theta$, in (\ref{eq_|S|}).
Feger et al. have proposed a least squares estimation (LSE) algorithm
that maximizes the function $J(r,u)$, using
a localized grid-search \cite{Feger2008}.
The function $J(r,u)$ is
essentially identical to $|S(x, y)|$ in (\ref{Sxy})
after the change of variables.
%
%which essentially
%maximizes
%in (\ref{S_1}) using a localized grid search~\cite{Feger2008}.
%This LSE algorithm
%
%for finding $r^\prime$ and $u^\prime$ that maximizes a function $J(r,u)$, using
%a grid-search
%\cite{Feger2008}.
%It can be seen that $J(r^\prime,u^\prime)$ is
%essentially identical to $|S(x^\prime, y^\prime)|$ in
%(\ref{S_1}), after the change of variables.
%considering the range difference per antenna
%by providing cost function as below.
% %
% \begin{gather}
%    J({r_k},{u_k}) = |\sum\limits_{n = 0}^{N - 1} {\sum\limits_{m = 0}^{M - 1}
%{z[n,m]{e^{ - j2\pi \frac{{{u_k}}}{\lambda }m}}{e^{ - j2\pi (2{r_k} + m{u_k})\frac{B}{{cN}}n}}} } |.
%\end{gather}
%%
%All sets of $(r_k,u_k)$ are determined where $J({r_k},{u_k})$ has a peak.
This LSE algorithm gives good estimation performance for a single target, because
it retains the range-angle coupling term.
However for a target with a completely unknown position, a full grid search is needed with
high computational cost.
Also, when multiple targets present,
the interference between targets causes error on the range and angle estimates.

%In Section \ref{sec:simulation}, we will check the interference errors on range and angle estimation.

%%%%%%%%%%%%%%%%%%%%%%%%%%%%%%%%%%%%%%%%%%%%%%%%%%%%%%%
% Section 4. Maximum Likelihood Approach
%%%%%%%%%%%%%%%%%%%%%%%%%%%%%%%%%%%%%%%%%%%%%%%%%%%%%%%
\section{Proposed estimation algorithm}
\label{sec:ConventionalApproach}
%\subsection{ Likelihood Function }
%
The objective is to find the maximum likelihood estimates of $\psi_k$, $a_k$, $u_k$ and $r_k$,
where $\psi_k$ and $a_k$ are nuisance parameters. Let us first rewrite the signal (\ref{s_nm}) as
\begin{align}
{s_k}[n,m] =
{a_k}{e^{j{\psi _k}}}{e^{j2\pi \frac{{{u_k}}}{\lambda }m}}{g_k}[n,m],
\end{align}
where
$g_k[n,m] = e^{j2\pi (2{r_k} + m{u_k})\frac{B}{{cN}}n}$.
We stack the time samples of measurement (\ref{eq_m}) into a column vector
%We convert it to vector form,
${{\bf{z}}^{(m)}} \in \mathbb{C}^{N \times 1}$, as below.
\begin{gather}
    {{\bf{z}}^{(m)}} = \sum\limits_{k = 1}^K {{\bf{s}}_{(k)}^{(m)}}  + {{\bf{w}}^{(m)}},
\end{gather}
where
\begin{gather}
    {\bf{s}}_{(k)}^{(m)} = {a_k}{e^{j{\psi _k}}}{e^{j2\pi \frac{{{u_k}}}{\lambda }m}}{\bf{g}}_{(k)}^{(m)},
    \label{eq_s_bold}
    \\
    {\bf{g}}_{(k)}^{(m)} = [g_k[0,m], g_k[1,m], ..., g_k[N-1,m]]^T,
    \\
    {{\bf{w}}^{(m)}} = {[w[0,m],w[1,m],...,w[N - 1,m]]^T}.
\end{gather}
%
%%%%%%%%%%%%%%%%%%%%%%%%%%%%%%%%%%%%
%
%The noise is assumed as spatially and temporarily uncorrelated Gaussian noise.
Then, the likelihood function is
{\small{
\begin{align}
    \Lambda_0 &= \sum\limits_{m=0}^{M-1} \left[ {\bf z}^{(m)} - \sum\limits_{k = 1}^K {\bf s}_{(k)}^{(m)} \right]^H
    \cdot \frac{1}{\sigma^2} I \cdot \left[ {\bf z}^{(m)} - \sum\limits_{k = 1}^K {\bf s}_{(k)}^{(m)} \right]
%    \nonumber
    \\
    & = \frac{1}{\sigma^2} \sum\limits_{m = 0}^{M - 1} \left[ {{\bf z}^{(m)}}^H \cdot {\bf z}^{(m)} \right]
    \nonumber
    %\label{eq_ml_z}
    \\
    & \hspace{0.3cm}
    - \frac{1}{\sigma^2} \sum\limits_{m = 0}^{M - 1}
    \left[ {{\bf{z}}^{(m)}}^H \cdot \sum\limits_{k = 1}^K {\bf s}_{(k)}^{(m)}
    + \left(\sum\limits_{k = 1}^K {\bf s}_{(k)}^{(m)} \right)^H \cdot {\bf z}^{(m)} \right]
    %\label{eq_ml_sz}
    \nonumber
    \\
    & \hspace{0.3cm}
     + \frac{1}{\sigma ^2} \sum\limits_{m = 0}^{M - 1}
     \left[ \left( \sum\limits_{k = 1}^K {\bf s}_{(k)}^{(m)} \right)^H
     \cdot \sum\limits_{k = 1}^K {\bf s}_{(k)}^{(m)} \right],
    \label{eq_ml_ss}
\end{align}
}}%
where the superscript $H$ denotes the conjugated transposition.
After inserting (\ref{eq_s_bold}) into (\ref{eq_ml_ss}) and
removing unnecessary terms, the likelihood function becomes
\begin{align}
& \Lambda ({a_1},{\psi _1},{r_1},{u_1},...,{a_K},{\psi _K},{r_K},{u_K})
\nonumber
\\
& = - \sum\limits_{m = 0}^{M - 1} \sum\limits_{k = 1}^K
{a_k} \left[{e^{j({\psi _k} + 2\pi \frac{{{u_k}}}{\lambda }m)}}R_{(k)}^{(m)} + {e^{ - j({\psi _k}
+ 2\pi \frac{{{u_k}}}{\lambda }m)}}R_{(k)}^{(m)*} \right]
\nonumber
\\
& \hspace{0.2cm} + \sum\limits_{m = 0}^{M - 1} {\sum\limits_{k = 1}^K {\sum\limits_{l \ne k}^K {{a_k}{a_l}{e^{j({\psi _k}
+ 2\pi \frac{{{u_k}}}{\lambda }m)}}{e^{ - j({\psi _l} + 2\pi \frac{{{u_l}}}{\lambda }m)}}R_{(l,k)}^{(m)}} } }
\nonumber
\\
& \hspace{0.2cm}
+ MN\sum\limits_{k = 1}^K {{a_k}^2} ,
\label{eq_lambda}
\end{align}
where
{\small{
\begin{align}
R_{(k)}^{(m)} & = {{\bf{z}}^{(m)H}} \cdot {\bf{g}}_{(k)}^{(m)}
%\nonumber
%\\
%& =
%=\sum\limits_{n = 0}^{N - 1} {{z^*}[n,m] {g_k}[n,m]}
%\nonumber
%\\
%& =
=\sum\limits_{n = 0}^{N - 1} {{z^*}[n,m] {e^{j2\pi (2{r_k} + m{u_k})\frac{B}{{cN}}n}}},
\label{rk}
\\
R_{(l,k)}^{(m)} & = {\bf{g}}_{(l)}^{(m)H} \cdot {\bf{g}}_{(k)}^{(m)}
%\\
%& =
%=\sum\limits_{n = 0}^{N - 1} {{g_l}^*[n,m]  {g_k}[n,m]}
%\\
%&
= \sum\limits_{n = 0}^{N - 1} {{e^{j2\pi (2{r_k} + m{u_k} - 2{r_l} - m{u_l})\frac{B}{{cN}}n}}}.
\label{rkl}
\end{align}
}}
Here, $R_{(l,k)}^{(m)}$ represents
interferences from other targets, and retaining this term allows the proposed algorithm to achieve
the CRB, even in multiple target environment.

%\subsection{ Parameters estimation }
%\label{sec:parameter_estimation}
%
The parameters, $\psi_k$, $a_k$, $u_k$, and $r_k$ can be found by maximizing $\Lambda( \cdot)$ in (\ref{eq_lambda}).
Because $\Lambda(\cdot)$ is nonlinear, however, $u_k$, $r_k$ and some other intermediate parameters
needed for $\psi_k$, $a_k$ should be found by Newton-Raphson iterations.
Detailed derivations for the formulas below are given in Appendix \ref{appendix_ML}.

First, we determine $\hat \psi_k$ for all $k$ by
\begin{gather}
    {\hat \psi_k} =  - \angle (\sum\limits_{m = 0}^{M - 1} {{e^{j2\pi \frac{{{u_k}}}{\lambda }m}}S_{(k)}^{(m)}} ),
%    \\
%    S_{(k)}^{(m)} = R_{(k)}^{(m)} - \sum\limits_{l \ne k}^K {{a_l}{e^{ - j({\psi _l} + 2\pi \frac{{{u_l}}}
%    {\lambda }m)}}R_{(l,k)}^{(m)}},
%    \label{eq_psi_k}
\end{gather}
where
$S_{(k)}^{(m)} = R_{(k)}^{(m)} - \sum\limits_{l \ne k}^K {{a_l}{e^{ - j({\psi _l} + 2\pi \frac{{{u_l}}}
{\lambda }m)}}R_{(l,k)}^{(m)}}$, and $\angle(\cdot)$ denotes the phase of ($\cdot$).
See Appendix A.1.

Second, $\hat a_k$ for all $k$ are found by solving the linear system of equations,
\begin{align}
    {\bf{B}} \cdot {\bf{a}} = {\bf{y}}, \quad {\bf{B}} \in \mathbb{R}^{K \times K}, \quad
{\bf{y}} \in \mathbb{R}^{K \times 1},
    \label{eq_a_k}
\end{align}
where
${\bf{a}} = {[{\hat  a_1},...,{\hat a_K}]^T} \in \mathbb{R}^{K \times 1} $ ,
%${\bf{B}} \in \mathbb{R}^{K \times K} $ ,
%and ${\bf{y}} \in \mathbb{R}^{K \times 1} $.
%The components of ${\bf{B}}$ and ${\bf{y}}$ are
and
\begin{align}
&{\bf{B}}(k,k) = MN, \quad 1 \le k \le K ,
\\
&{\bf{B}}(k,l) = \sum\limits_{m = 0}^{M - 1} {{\mathop{\rm Re}\nolimits} [{e^{j({\psi _k} + 2\pi \frac{{{u_k}}}
{\lambda }m)}}{e^{ - j({\psi _l} + 2\pi \frac{{{u_l}}}{\lambda }m)}}R_{(l,k)}^{(m)}]},
\nonumber
\\
& \hspace{2cm}
1 \le k,l \le K, \quad k \ne l ,
\\
&{\bf{y}}(k) = \sum\limits_{m = 0}^{M - 1} {{\mathop{\rm Re}\nolimits} [{e^{j({\psi _k}
+ 2\pi \frac{{{u_k}}}{\lambda }m)}}R_{(k)}^{(m)}]}, \quad 1 \le k \le K.
\end{align}
%where
%$k=1,...,K$, $l=1,...,K$, and $l \ne k$.
See Appendix A.2.

Third, $\hat u_k$ for all $k$ are determined iteratively as
\begin{gather}
    {\hat u_k}(i + 1) = {\hat u_k}(i) - \frac{f_u}{{f_u'}}{|_{{\hat u_k} = {\hat u_k}(i)}},
\end{gather}
where $\hat u_k(i)$ denotes $\hat u_k$ at the $i$th iteration, and
$f_u$, $f_u^\prime$ are derived in Appendix A.3.
%The $f_u$ and $f_u'$ are expressed as

Fourth, we determine $\hat r_k$ iteratively as
\begin{gather}
    {\hat r_k}(i + 1) = {\hat r_k}(i) - \frac{f_r}{{f_r'}}{|_{{\hat r_k} = {\hat r_k}(i)}},
\end{gather}
where $\hat r_k(i)$ denotes $\hat r_k$ at the $i$th iteration, and
$f_r$, $f_r^\prime$ are derived in Appendix A.4.
%
%The $f_r$ and $f_r'$ are expressed as
%
%\begin{align}
%& f_r =  - 2{a_k}\sum\limits_{m = 0}^{M - 1} {} {\mathop{\rm Re}\nolimits} [{e^{j({\psi _k} + 2\pi \frac{{{u_k}}}
%{\lambda }m)}}\frac{{\partial S_{(k)}^{(m)}}}{{\partial {r_k}}}],
%\\
%& f_r' =  - 2{a_k}\sum\limits_{m = 0}^{M - 1} {} {\mathop{\rm Re}\nolimits} [{e^{j({\psi _k} + 2\pi
%\frac{{{u_k}}}{\lambda }m)}}\frac{{{\partial ^2}S_{(k)}^{(m)}}}{{\partial {r_k}^2}}],
%\end{align}
%where
%$\frac{{\partial S_{(k)}^{(m)}}}{{\partial {r_k}}}$ and $\frac{{{\partial ^2}S_{(k)}^{(m)}}}
%{{\partial {r_k}^2}}$ are given in (\ref{eq_S_r}) and (\ref{eq_S_r2}).
%
%In the above, the Jacobians $f_u$, $f_r$ and the Hessians $f_u^\prime$, $f_r^\prime$ are derived in
%Appendix \ref{appendix_ML}.
%
%\subsection{ Initialization and iterative process }

The proposed joint range and angle estimation algorithm, including the initialization and termination,
is summarized as Algorithm \ref{algorithm}.
%For initialization, we determine $u_k(0)$ and $r_k(0)$ by performing $(N \times M)$ grid search
%on $J(r_k,u_k)$ described in Section \ref{sec:LSE}.
%To estimate $\psi_k(0)$, we use $R_{(k)}^{(m)}$ instead of $S_{(k)}^{(m)}$ in (\ref{eq_psi_k}),
%which technically means that the interference term, $R_{(l,k)}^{(m)}$, is ignored in (\ref{eq_S_psi}) for initial value finding.
%We solve the linear system, (\ref{eq_a_k}), to estimate $a_k(0)$.
%Then, we estimate all the parameters iteratively, as described in Section \ref{sec:parameter_estimation},
%until $\Lambda$ converges.

%%%%%%%%% Algorithm
 \begin{algorithm}[H]
 \caption{Algorithm for joint range and angle estimation}
 \label{algorithm}

 % 1. Initialization
 \textbf{Initialization:}
% Do 2D-FFT process
 \begin{algorithmic}[1]
  \STATE Determine $ {u_k}(0)$ and ${r_k}(0)$ by $N \times M$
  grid search on $|S(x^\prime, y^\prime)|$.
  \STATE Determine ${\psi _k}(0) =  - \angle (\sum\limits_{m = 0}^{M - 1}
  {{e^{j2\pi \frac{{{u_k}}}{\lambda }m}}R_{(k)}^{(m)}})$ using (\ref{rk}).
  \STATE Determine ${a_k}(0)$ by solving (\ref{eq_a_k}).
 \end{algorithmic}

 % 2. Loop
  \textbf{Iteration:}
  \begin{algorithmic}[1]
  \STATE $i=0$;
  \REPEAT
%  \FOR {$i = 0$ to $N_i-1$}
  \STATE Compute $R_{(k)}^{(m)},
   R_{(l,k)}^{(m)},
   S_{(k)}^{(m)},
  \frac{{{\partial}S_{(k)}^{(m)}}}{{\partial {u_k}}},
  \frac{{{\partial ^2}S_{(k)}^{(m)}}}{{\partial {u_k}^2}},
  \frac{{\partial S_{(k)}^{(m)}}}{{\partial {r_k}}},
  \frac{{{\partial ^2}S_{(k)}^{(m)}}}{{\partial {r_k}^2}} $
  using $u_k(i)$ and $r_k(i)$
  \STATE Update ${\psi _k}(i + 1)$, ${a_k}(i + 1)$, ${u _k}(i + 1)$, and ${r_k}(i + 1)$
  for all $k$
  \STATE $i=i+1$

  \UNTIL { $|\Lambda (i) - \Lambda (i-1)| < \delta$ }

 %\RETURN ${u _k ({\rm{end}})}$ and ${r _k ({\rm{end}})}$ for all $k$
 \end{algorithmic}

 \end{algorithm}
%%%%%%%%%

%%%%%%%%%%%%%%%%%%%%%%%%%%%%%%%%%%%%%%%%%%%%%%%%%%%%%%%
% Section 5. Simulation Results
%%%%%%%%%%%%%%%%%%%%%%%%%%%%%%%%%%%%%%%%%%%%%%%%%%%%%%%
\section{Performance analysis}
\label{sec:simulation}

%\subsection{ Parameters }

For simulation study, the chosen operating frequency, bandwidth and the sweep time are, respectively,
$f_c = 77$ GHz, $B = 4$ GHz, and $S=10^{-4}$ s, so that the true range resolution is
$\Delta r = c/(2B) = 3.75$ cm.
FMCW radar chips of similar specifications are now appearing in the
commercial market. In addition, the number of range samples, number of transmitting antennas and number of
receive antennas are assumed to be $N=256$, $P=4$, and $Q=4$, respectively, so that the number of virtual receive
antennas is $M=PQ=16$.

We apply a large amount of zero-padding for the 2D-FFT, and an extremely fine search,
for the 2D-MUSIC and LSE,
by the factor of 2048,
both to the range and angle dimension.
This will ensure that the grid-based algorithms; 2D-FFT, 2D-MUSIC and LSE are fairly compared with the
proposed gridless algorithm, even though the computational cost of this zero-padding is prohibitively high.
With this oversampling,
the apparent range and angle resolutions are, respectively $\Delta r/2048 = 1.83 \times 10^{-5}$m and
$\Delta \theta /2048 = \Delta u /(du/d\theta) /2048 = 2/(M \times {\rm{cos}}(15^o)) \times 180/ \pi /2048 = 0.0036^o$,
where the angle resolution is calculated at $\theta = 15 ^o$ on which the target is assumed to be located.
%The half of the resolution is $9.15 \times 10^{-6} (m)$ and $1.8 \times 10^{-3} (^o)$.
If the root-mean squared errors (RMSEs) of these grid-based algorithms
are larger than the half of the apparent resolution, which are
$9.15 \times 10^{-6}$ m for the range, and $0.0018^o$ for the angle, then
the error is not caused
by the lack of resolution, but by the bias of these grid-based algorithms.

%%%%%%%%%%%%%%%%%%%%%%
\subsection{Single target }

A target is assumed to be located at $r = 5$ m, and $\theta = 15^o$.
%the range bin difference per antenna is $ \delta r = 0.0067 (bin) $ according to (\ref{bin_dif}).
Then the true range bin index and the true angle bin index, allowing fractions are, respectively
$n_r = r/{\Delta r} = 133.333 $, and $ m_{\theta} = M/2 \times \sin (15 ^o) = 2.071 $.

\begin{figure}[!h]
\centering
% sub 1
\begin{subfigure}{.5\textwidth}
  \centering
  \includegraphics[width=\columnwidth]{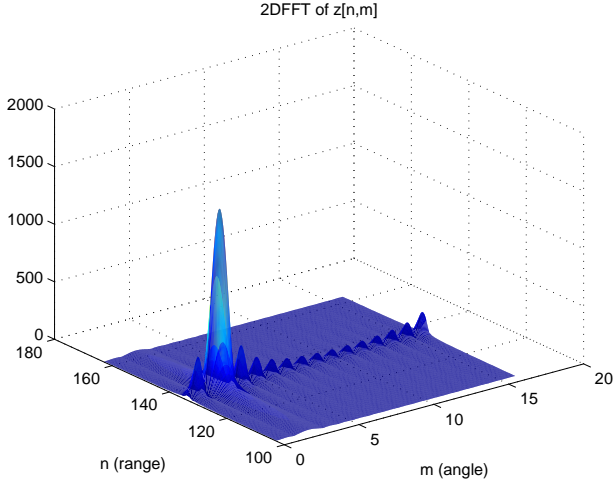}
  \caption{}
  \label{sim_1Tg_c1_2DFFT}
\end{subfigure}%
\hfil
% sub 2
\begin{subfigure}{.5\textwidth}
  \centering
  \includegraphics[width=\columnwidth]{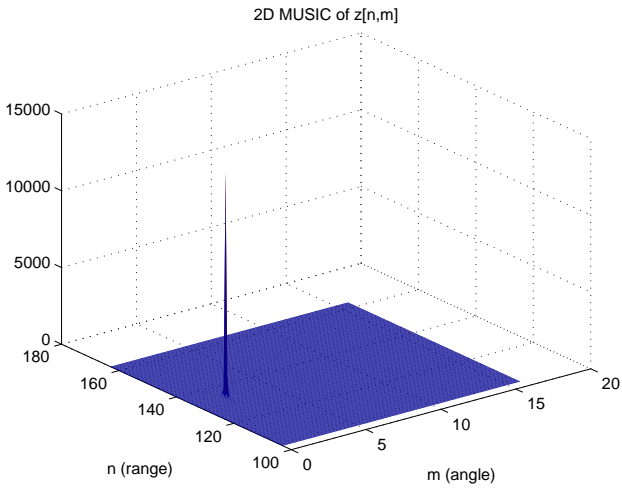}
  \caption{}
  \label{sim_1Tg_c1_2DMUSIC}
\end{subfigure}
\caption{
The 2D frequency estimation algorithm applied to the noiseless measurement of a single target gives
bias error both in the range and angle.
%Single target with small range bin difference, $\delta r = 0.0067$.
%The 2D frequency estimation method such as 2D-FT and 2D-MUSIC causes bias error on range and angle estimation.
(a) 2D-FFT spectrum.
(b) 2D-MUSIC spectrum using the spatial smoothing with a $10 \times 10$ sub-matrix.
}
\label{sim_1Tg_c1_cost}
\end{figure}

% Fig: 2D-FFT and 2D-MUSIC cost fuction
Fig. \ref{sim_1Tg_c1_cost} shows the spectra of the 2D-FFT
and 2D-MUSIC with noiseless $z[n,m]$, where the 2D-MUSIC exploits
the spatial smoothing with a $ 10 \times 10 $ sub-matrix \cite{Belfiori2012a}.
For both 2D-FFT and 2D-MUSIC, the peak bin indices are $n_p = 133.383$ and $m_p = 2.124$, and therefore
corresponding range and angle estimates are respectively, $5.00186$ m and $15.397^o$.
The bias errors of these estimates are $0.00186$ m and $0.397^o$, which are
almost the same to the theoretical values,
$\Delta {r^{(b)}} = 0.0019$ m and
$\Delta {\theta ^{(b)}} = {0.399^o}$ obtained from (\ref{eq_r'b}) and (\ref{eq_theta'b}).

\begin{figure}[!h]
\centering
% sub 1
\begin{subfigure}{.5\textwidth}
  \centering
  \includegraphics[width=\columnwidth]{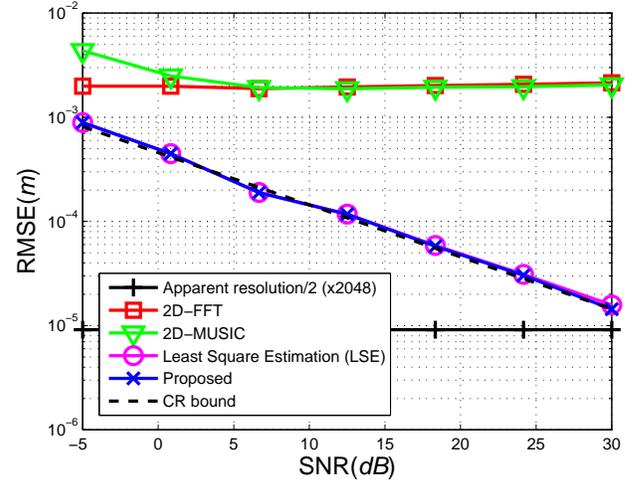}
  \caption{}
  \label{sim_1Tg_c1_range}
\end{subfigure}%
\hfil
% sub 2
\begin{subfigure}{.5\textwidth}
  \centering
  \includegraphics[width=\columnwidth]{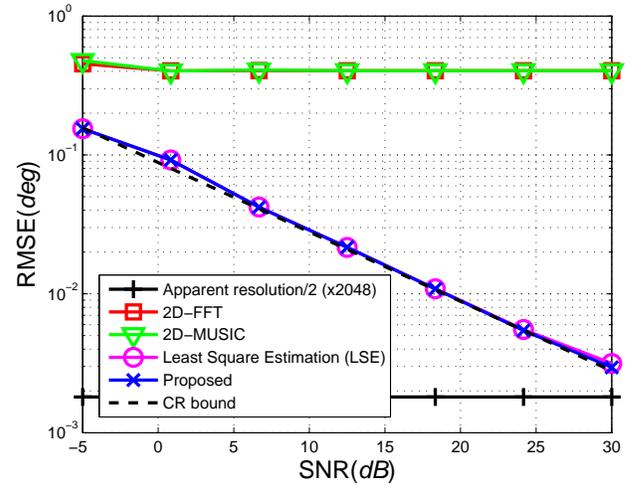}
  \caption{}
  \label{sim_1Tg_c1_angle}
\end{subfigure}
\caption{
%Single target with small range bin difference, $\delta r = 0.0067$.
RMSEs for the single target. %of range and angle estimation versus SNR.
(a) The range estimates.
(b) The angle estimates.
}
\label{sim_1Tg_c1_RMSE}
\end{figure}

% Fig: RMSE
Fig. \ref{sim_1Tg_c1_RMSE} shows the RMSEs of range and angle estimates
against signal to noise ratio (SNR) defined as $Pa^2/\sigma^2$.
%In (\subref{sim_1Tg_c1_range}) and (\subref{sim_1Tg_c1_angle}), the RMSEs of range and angle are described, respectively.
The simulation is a results of 300 trials using uniformly distributed
$\psi \sim U(0,2\pi)$.
Note that both for 2D-FFT and 2D-MUSIC, the RMSEs are much larger than the apparent resolution,
$1.83 \times 10^{-5}$ m and $0.0036^o$,
and therefore the
errors are the bias errors, not the resolution errors.
On the other hand, the LSE and the proposed algorithm give the RMSEs which coincide with the CRBs.
The CRB is derived in Appendix \ref{sec:CRB}.
%
%
%With LSE, we determine $r$ and $\theta$ to maximize $J(r,u)$  based on ($\times 2048$) grid search on range and angle domain.
%The RMSE is the same as CRB since a grid search was performed densely enough.
%
%With proposed method, the RMSE along SNR is the same as CRB.

% Fig of range,angle performance
%%%%%%%%%%%%%%%%%
%%%%%%%%%%%%%%%%%

% Fig of range,angle performance
%%%%%%%%%%%%%%%%%
%%%%%%%%%%%%%%%%%

%%%%%%%%%%%%%%%%%%%%%%
\subsection{Multiple targets $(K=2)$}
%%%%%%%%%%%%%%%%%%%%%%

In addition to the target at $r_1=5$ m and $\theta_1 = 15^o$, we add another target at
$r_2 = 5$ m and $\theta_2 = -15^o$.
The true range bin indices of the two targets are the same,
$ n_ {r1} = n_ {r2} = r_1 / {\Delta r} = 133.333 $,
while the true angle bin indices
are $ m_{\theta_1} = M/2 \times \sin(15^o) = 2.071 $ and $ m_{\theta_2} = M/2 \times \sin (-15 ^o) + M = 13.929 $.

\begin{figure*}[!h]
\centering
% sub 1
\begin{subfigure}{.33\textwidth}
  \centering
  \includegraphics[width=\columnwidth]{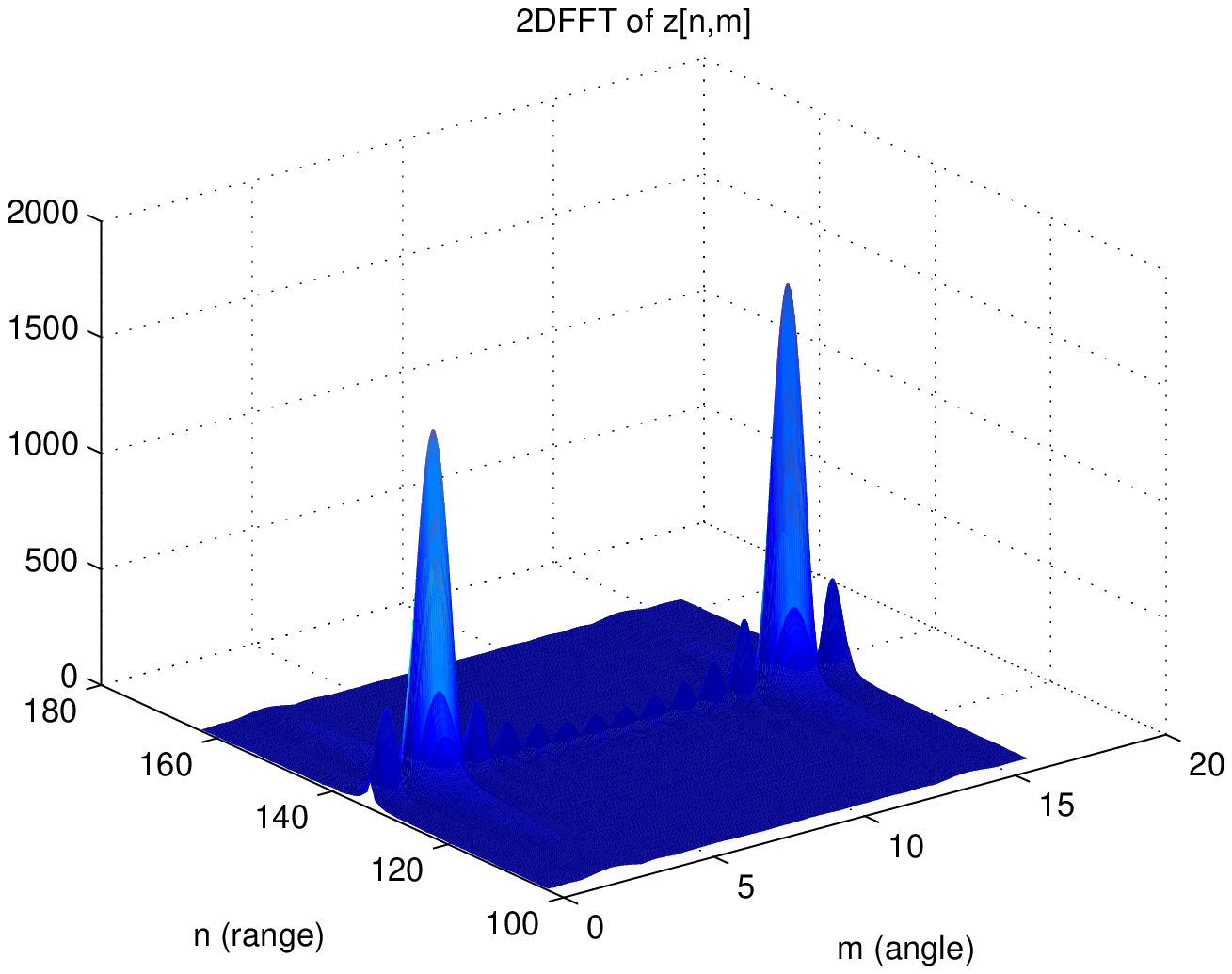}
  \caption{}
  \label{sim_2Tg_c1_2DFFT}
\end{subfigure}%
%\hfil
% sub 2
\begin{subfigure}{.33\textwidth}
  \centering
  \includegraphics[width=\columnwidth]{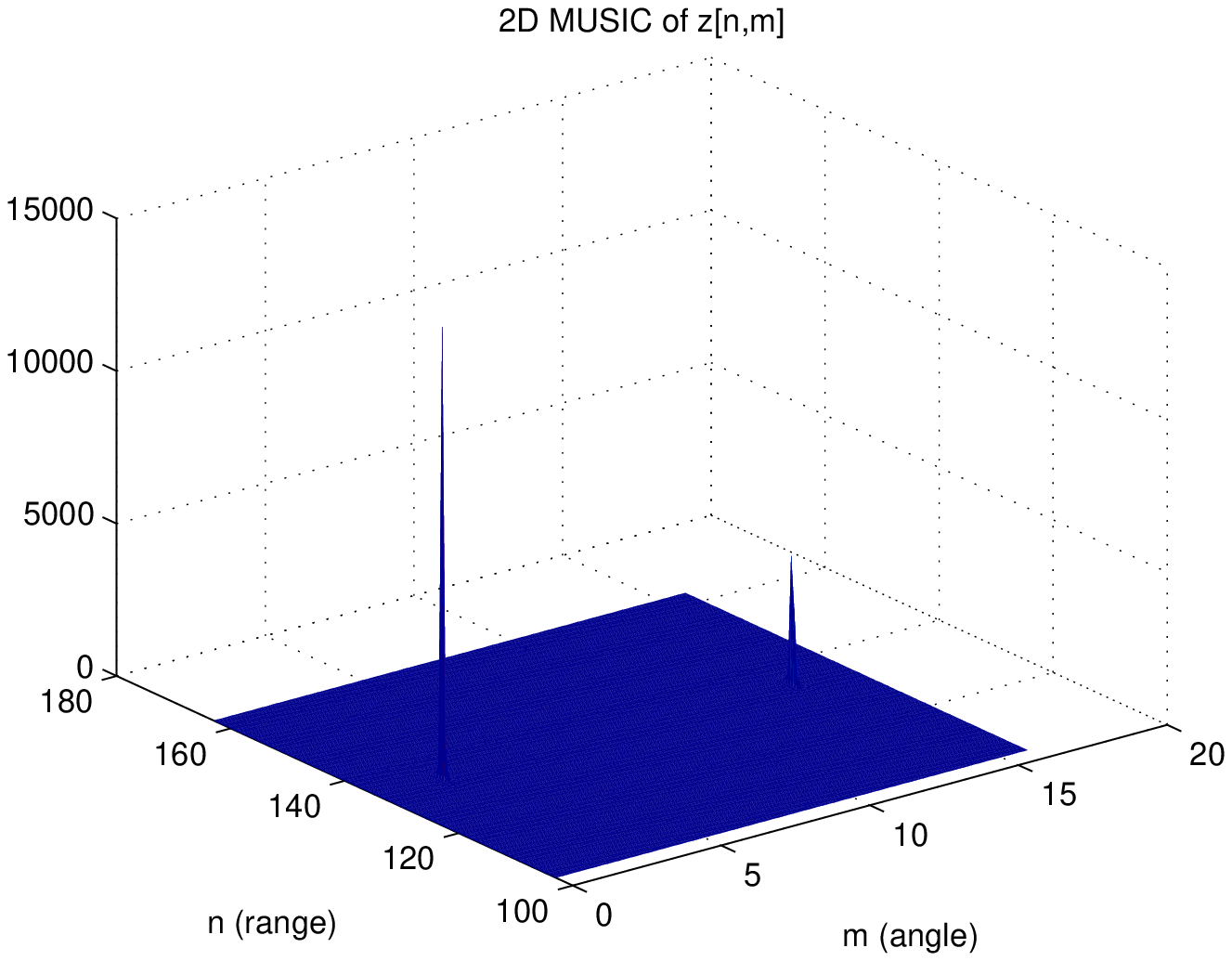}
  \caption{}
  \label{sim_2Tg_c1_2DMUSIC}
\end{subfigure}
% sub 3
\begin{subfigure}{.33\textwidth}
  \centering
  \includegraphics[width=\columnwidth]{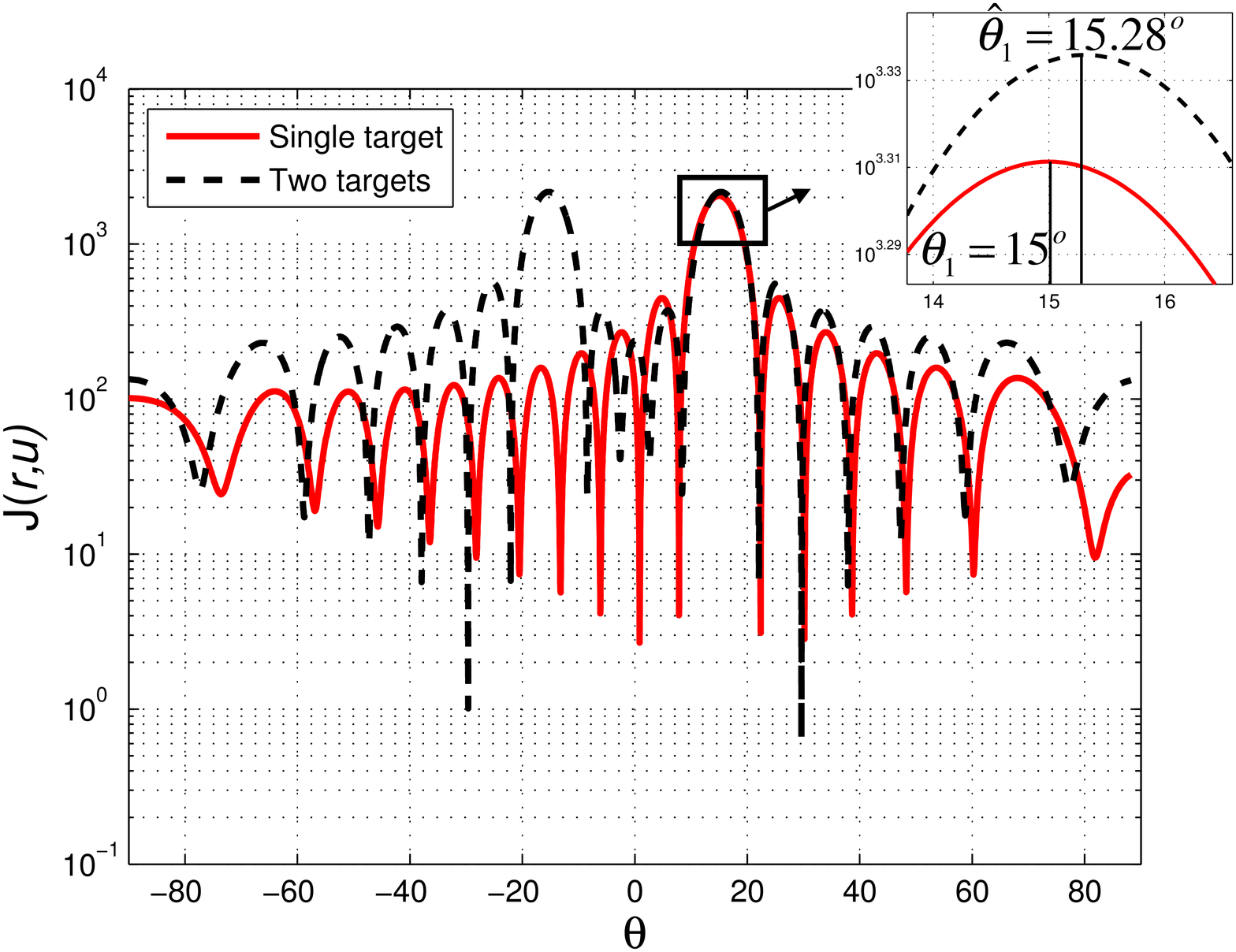}
  \caption{}
  \label{sim_2Tg_c1_LSE}
\end{subfigure}
\caption{
%Two targets with small range bin difference, $\delta r_1 = 0.0067$ and $\delta r_2 = -0.0067$.
%In range and angle estimation,
For noiseless measurement of two targets, the 2D-FFT and 2D-MUSIC algorithms give bias error plus interference error,
while the LSE algorithm gives interference error
%
%The range and angle estimates of two targets.
(a) The 2D-FFT spectrum.
(b) The 2D-MUSIC spectrum using spatial smoothing with a $10 \times 10$ sub-matrix.
(c) The LSE function, $J(r,u)$ at $r=5$m.
%(a) 2D-FT of $z[n,m]$.
%(b) 2D-MUSIC spectrum with spatial smoothing technique with $10 \times 10$ sub-matrix.
%(c) $J(r,u)$ of LSE on the intersection of $r=5(m)$.
}
\label{sim_2Tg_c1_cost}
\end{figure*}

% Fig 3
Fig. \ref{sim_2Tg_c1_cost}
shows the 2D-FFT, 2D-MUSIC spectra and the LSE function,
$J(r,u)$ at the intersection of $r = 5$ m, all for noiseless $z[n,m]$.
In Fig. \ref{sim_2Tg_c1_cost}(\subref{sim_2Tg_c1_2DFFT}), the indices $[n_{p}, m_{p}]$ of the two peaks are [133.383, 2.131] and
[133.383, 13.869], which correspond to the range and angle estimates of
[$5.00186$ m, $15.448^o$] and
[$5.00186$ m, $-15.448^o$], respectively.
%and $[5.00186 (m), -15.448 (^o)]$,
%respectively.
The errors $0.00186$ m and $0.448^o$ %when estimated by 2D-FFT.
come from two sources, the bias error given in (\ref{eq_r'b})
and (\ref{eq_theta'b}), and the interference error between the two target responses.
In Fig. \ref{sim_2Tg_c1_cost}(\subref{sim_2Tg_c1_2DMUSIC}), the indices of the peaks %$[n_{p}, m_{p}]$
are [133.383, 2.131] and
[133.383, 13.876], which give the range and angle estimates,
[$5.00186$ m, $15.449^o$] and
[$5.00186$ m, $-15.397^o$], respectively.
%
%In Fig. \ref{sim_2Tg_c1_cost} (\subref{sim_2Tg_c1_2DMUSIC}), the peak indices in 2D-MUSIC are $ n_{p1} = 133.383 $,
%$ m_{p1} = 2.131 $ and $ n_{p1} = 133.383 $ and $ m_{p1} = 13.876 $.
%Then, the estimated range and angle is $[5.00186 (m), 15.397 (^ o)]$ and $[5.00186 (m), -15.397 (^o)]$ , respectively.
%
%The errors $0.00186$ m and $0.397^o$ are almost the same to theoretical values
%predicted by (\ref{eq_r'b})
%and (\ref{eq_theta'b}).
%Therefore, the 2D-MUSIC method is not affected by the interference from the other target.
%
As with the 2D-FFT case, the errors $0.00186$ m and $0.449^o$ of the first target (and $0.00185$ m
and $0.397^o$ for the second target) are a sum of the bias error and the interference error.
In Fig. \ref{sim_2Tg_c1_cost}(\subref{sim_2Tg_c1_LSE}), the LSE function $J(r,u)$ is plotted against
$\theta$, at the fixed $r=5$ m.
Note that the peaks occur at $15.28^o$ and $-15.28^o$,
while true angles are $\theta_1 = 15^o$ and $\theta_2 = -15^o$.
This large error is caused by the sidelobe of each other.
%from other targets interferes with the desired target.

% Fig of range,angle performance
%%%%%%%%%%%%%%%%%
%%%%%%%%%%%%%%%%%

\begin{figure*}[!h]
\centering
% sub 1
\begin{subfigure}{.5\textwidth}
  \centering
  \includegraphics[width=\columnwidth]{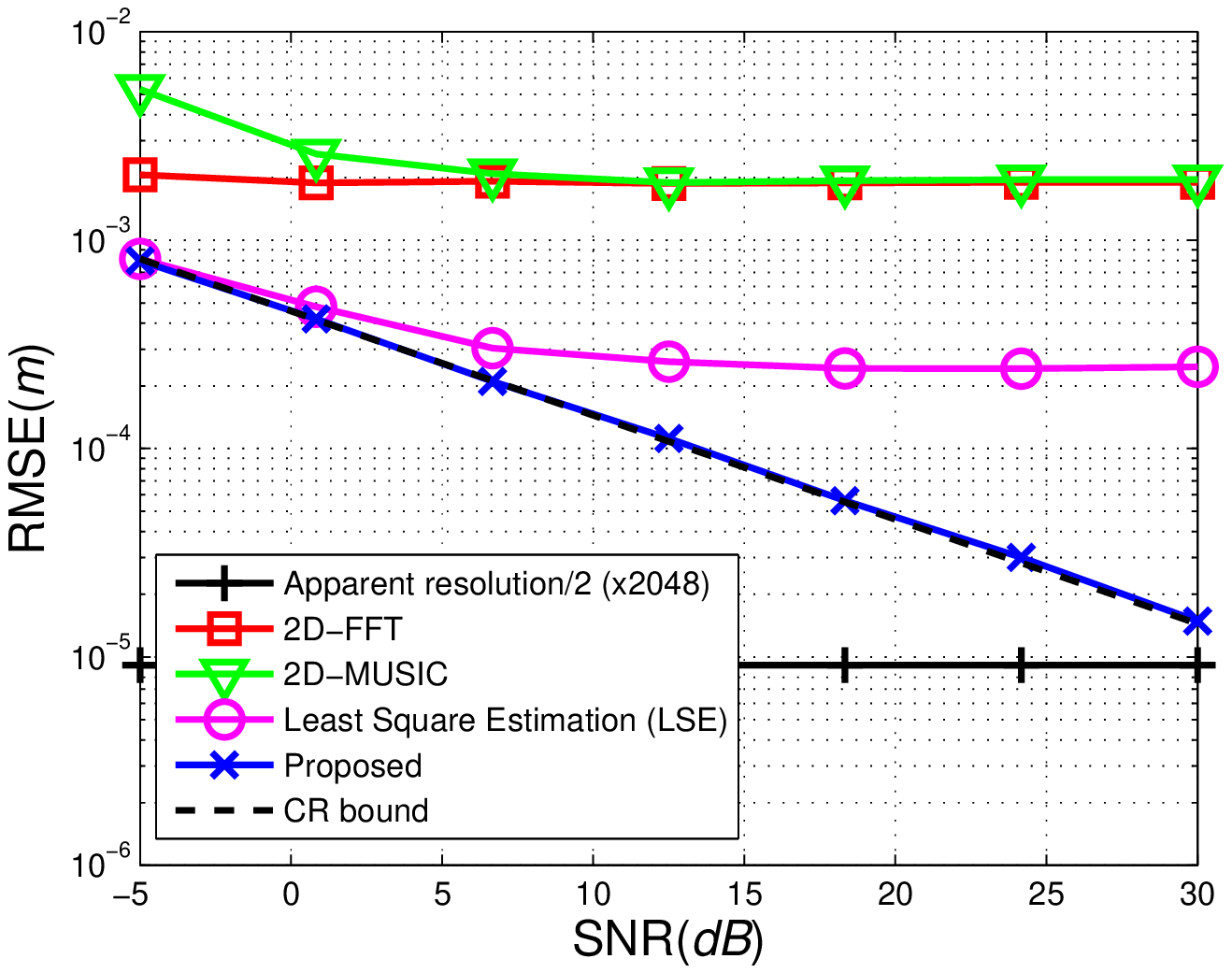}
  \caption{}
  \label{sim_2Tg_c1_range1}
\end{subfigure}%
%\hfil
% sub 2
\begin{subfigure}{.5\textwidth}
  \centering
  \includegraphics[width=\columnwidth]{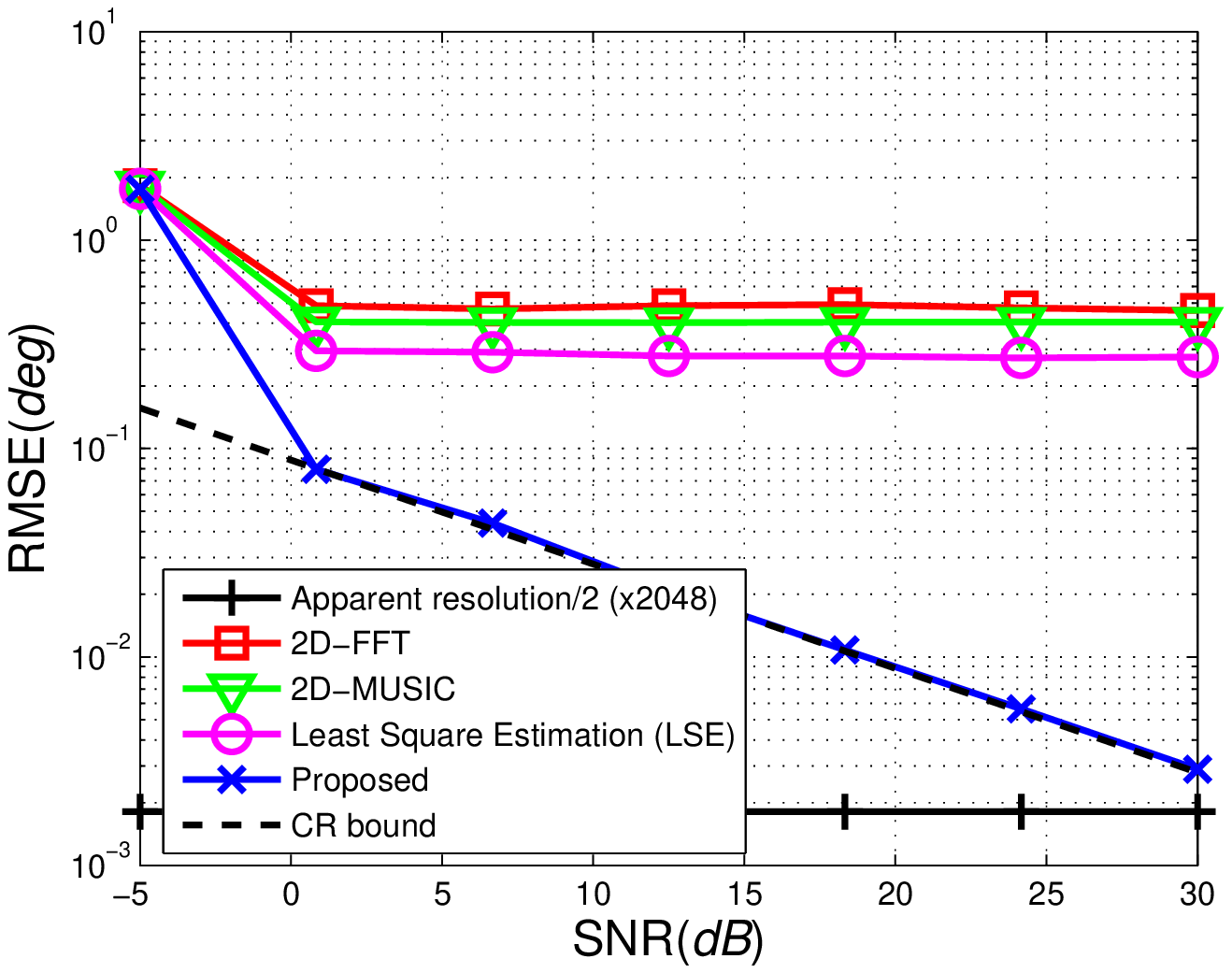}
  \caption{}
  \label{sim_2Tg_c1_angle1}
\end{subfigure}
\caption{
RMSEs for the multiple targets.
(a) The range estimates for the first target.
(b) The angle estimates for the first target.
The plots for the second target are almost identical to the first target.
}
\label{sim_2Tg_c1_RMSE}
\end{figure*}

% Fig : RMSE
In Fig. \ref{sim_2Tg_c1_RMSE}, the RMSE of the range estimate and the RMSE of the angle estimate
for the first target are depicted against the SNR.
Those plots for the second target are almost identical to the first target, and are omitted to conserve space.
With the 2D-FFT algorithm,
%with ($\times 2048$) zero-paddings on range and angle domain,
the range and angle estimation errors are $0.0019$ m and $0.45^o$, respectively for both targets.
With the 2D-MUSIC approach,
the estimates give errors of $0.0019$ m and $0.4^o$ for both targets.
Note that the proposed algorithm, which does not have bias or interference errors,
achieves the CRB, while
the other three grid-based algorithms, 2D-FFT, 2D-MUSIC and LSE do not.
%
%
% Fig of Tg2
%%%%%%%%%%%%%%%%%
%%%%%%%%%%%%%%%%%

%%%%%%%%%%%%%%%%%%%%%%%%%%%%%%%%%%%%%%%%%%%%%%%%%%%%%%%
% Parking Simulation
%%%%%%%%%%%%%%%%%%%%%%%%%%%%%%%%%%%%%%%%%%%%%%%%%%%%%%%
\section {Application to the autonomous valet parking}

%%%%%%%%%%%%%%%%%%%%%%%%%%%%%%%%%%%%%%%%%%%%%%%%%%%%%%%
\subsection{Estimation error in the scatterer positions}

\begin{figure}[!h]
\centering
% sub 1
\begin{subfigure}{.5\textwidth}
  \centering
  \includegraphics[width=\columnwidth]{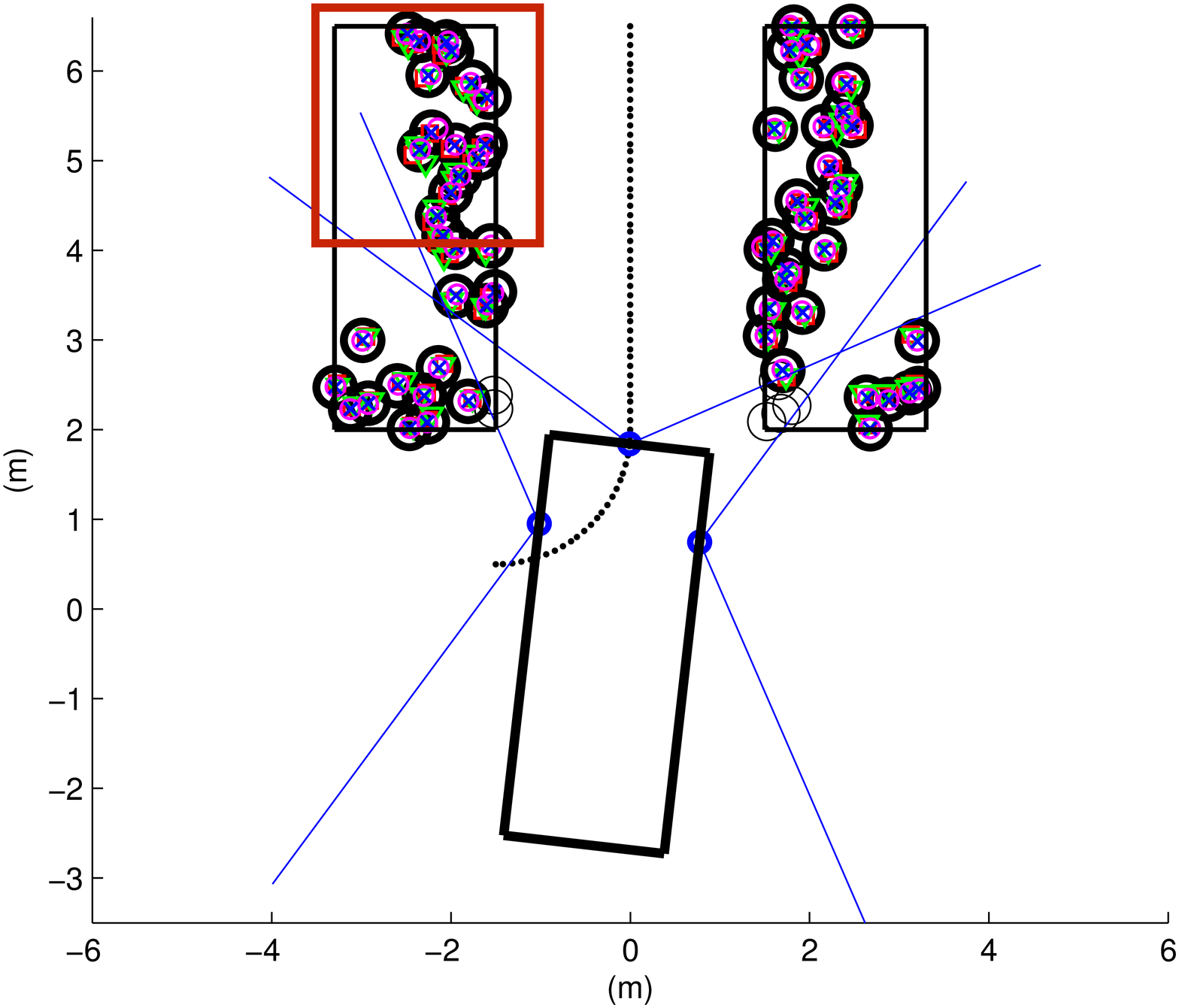}
  \caption{}
  \label{sim_PointCloud}
\end{subfigure}%
\hfil
% sub 2
\begin{subfigure}{.5\textwidth}
  \centering
  \includegraphics[width=\columnwidth]{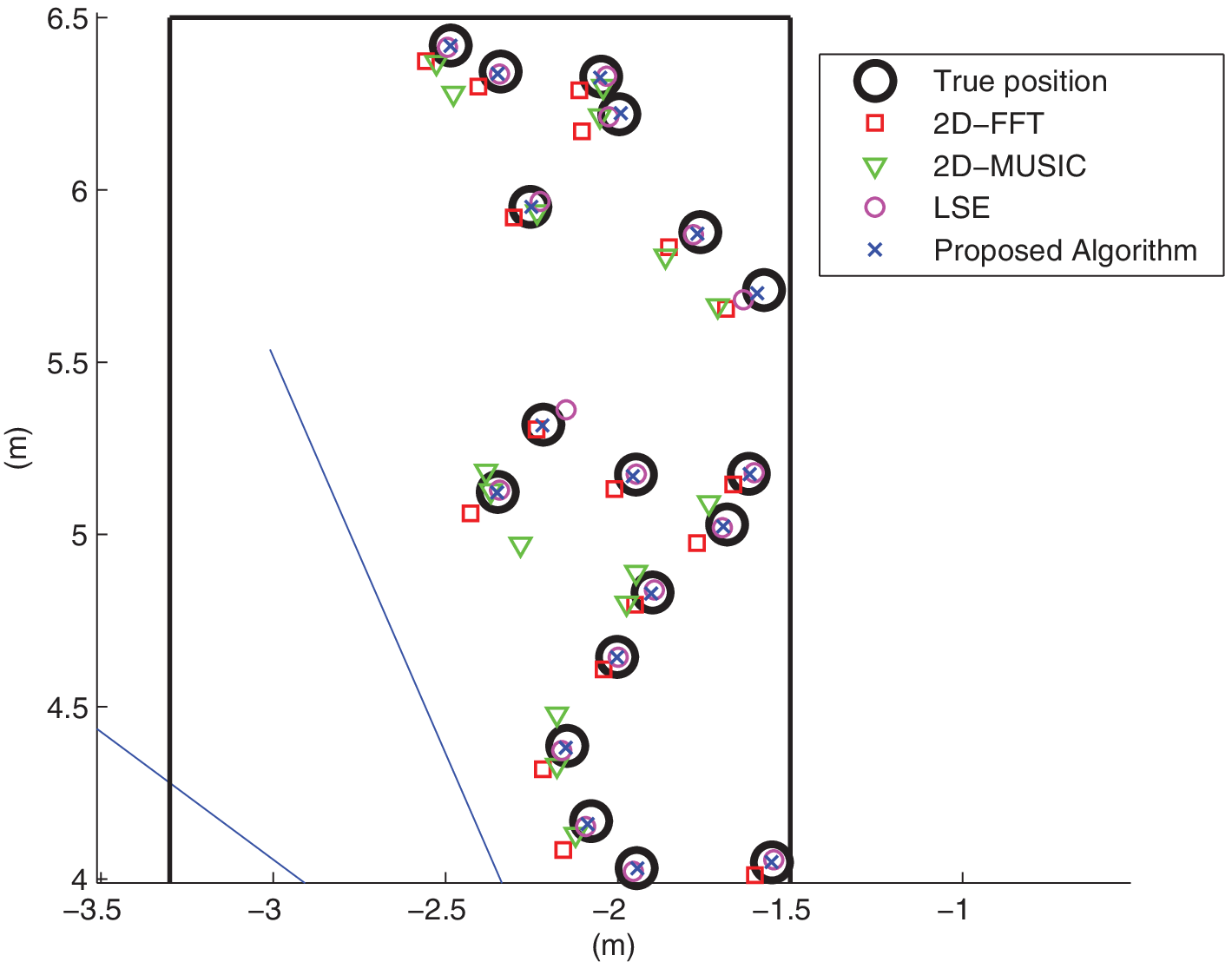}
  \caption{}
  \label{sim_PointCloudMag}
\end{subfigure}
\caption{
(a) The vehicle in the middle has three FMCW MIMO radars, which estimate the positions of the scatterers.
All three vehicles are stationary.
%At a specific location of the vehicle, three FMCW MIMO radar systems estimate the location of the point cloud.
(b) An enlarged view of the boxed area in (a).
}
\label{sim_PC}
\end{figure}

Fig. \ref{sim_PC}(\subref{sim_PointCloud}) shows three vehicles modeled as 2D rectangles;
two vehicles are parked and the other one is about to back into the slot in-between.
It is assumed that the size of the vehicle is $1.8$ m wide and $4.6$ m long,
and that the back-parking vehicle is
%
%We will compare the estimation performance of the estimators when the vehicle is located at
%a certain location shown in Fig. \ref{sim_PC} (\subref{sim_PointCloud}).
%The size of the car is $1.8$ m $\times$ $4.6$ m, which is a standard sedan size.
%, and the parking lot size is $2.3 (m) \times 5 (m)$.
%The car is
equipped with three radars, one on the rear and two on each side.
Each radar has $4\times4$ MIMO antennas of $120^o$ field of views (FOVs),
constituting a virtual array of sixteen antennas.
%where each radar has the field of view (FOV) of $120 (^o)$.
%Recall that
%$f_c = 77$ GHz, $B = 4$ GHz, and $\Delta r = 3.75$ cm.
%
It is also assumed that two parked vehicles give rise to 66 scatterers total,
which are marked with circles.
%The location of the point cloud is estimated by three FMCW MIMO radar systems on the car.
%
Fig. \ref{sim_PC}(\subref{sim_PointCloudMag})
shows an enlarged view of the upper-left boxed area of Fig. \ref{sim_PC}(\subref{sim_PointCloud}).
It is apparent that the proposed algorithm gives more accurate estimates than the
other three algorithms.
This is indeed the case as can be seen in TABLE \ref{table:RMSEPC}, which shows
%To be more specific, in TABLE \ref{table:RMSEPC},
the RMSEs of the range, angle and the position estimates of the scatterers,
averaged over 61 detected scatterers out of 66 scatterers.
Notice that the estimates of the proposed algorithm are better than the other three, by an order of magnitude.
%
%in each method are depicted.
%Among 66 scatters, 61 were visible by the radar FOV and all 61 were detected.
% by constant threshold method.
%The detection was performed using the constant false alarm rate (CFAR) technique.
%The RMSE was calculated for all 61 scatterers.
%In the RMSE of 2D position, 2D-MUSIC method was the largest at $0.116$ m, and the proposed method was
%the smallest at $0.008$ m.
%The angle error has a great influence on the 2D position error.
%We can confirm the location of the scatterers estimated by each method.
%
%
%The two dimensional (2D) estimated location error with 2D-FFT or 2D-MUSIC is larger than those with other methods.
%
% Fig of Point Cloud at Certain location
%%%%%%%%%%%%%%%%%
%%%%%%%%%%%%%%%%%
%
%To be more specific, in TABLE \ref{table:RMSEPC}, the RMSEs of the range, angle, and 2D position of
%the scatterers in each method are depicted.
%Among 66 scatters, 61 were visible by the radar FOV and all 61 were detected.
% by constant threshold method.
%The detection was performed using the constant false alarm rate (CFAR) technique.
%The RMSE was calculated for all 61 scatterers.
%In the RMSE of 2D position, 2D-MUSIC method was the largest at $0.116$ m, and the proposed method was the smallest at $0.008$ m.
%The angle error has a great influence on the 2D position error.
%
%For 2D-FFT and , the RMSE of angle is $1.31 (^o)$, which is larger than the other methods.
%We note that the angle error has a great influence on the 2D position error.
%
% Table
\begin{table*}[h]
\caption {
%At a certain location of vehicle, the comparison of RMSE over all detected scatterers in single frame.
Comparison of the RMSE for the position estimates in Fig 5.
%Radar 1,2,3 represents the rear, right, and left one in order.
}
\label{table:RMSEPC}
\centering
\begin{tabular}{|l|c|c|c|c|}
\hline
                                  & \multicolumn{1}{l|}{2D-FFT} & {2D-MUSIC} & LSE    & Proposed                        \\ \hline
%\# of scatterers inside FOV (radar 1,2,3) & \multicolumn{4}{c|}{61(45+10+6)}                                       \\ \hline
%\# of scatterers detected (radar 1,2,3)   & \multicolumn{4}{c|}{61(45+10+6)}                                       \\ \hline
RMSE of the range (m)          & 0.0047  & 0.017   & 0.0019  & 0.0006 \\ \hline
RMSE of the angle ($^o$)         & 1.30     & 1.99     & 0.40    & 0.13                            \\ \hline
RMSE of the 2D position (m)    & 0.070   & 0.116   & 0.025   & 0.008                          \\ \hline
\end{tabular}
\end{table*}

\subsection{Localization of the vehicle}
\label{subsec:VehiclePositionUpdate_ICP}

\begin{figure}[!h]
\centering
\includegraphics[width=\columnwidth]{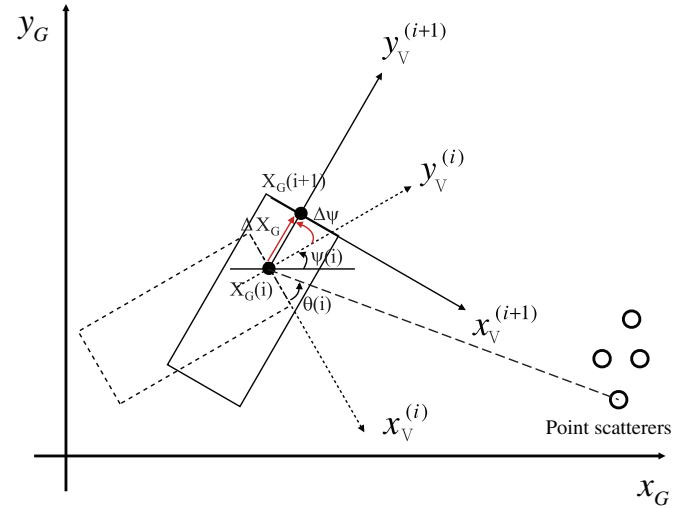}
\caption{
%The vehicle rotates by $\Delta \psi$ and moves forward by $\Delta {\bf{X}}_G$,
%of which values are calculated through iterative closest point (ICP) algorithm.
Geometry of the ground coordinate system and the vehicle coordinate systems at time $i$ and $i+1$.
%by comparing
%the relative position of point cloud seen from the vehicle over the frame.
%the change of point cloud position with respect to car coordinate by using iterative closest point (ICP) algorithm.
}
\label{fig_ICP}
\end{figure}

The range and angle estimates of the scatterers
can be used to estimate the position %${\bf{X}}_G(i)$
and the orientation angle
%$\psi (i)$
of the moving own-vehicle, which carries the radars, using the ICP method~\cite{Besl1992}
as explained below.
In Fig \ref{fig_ICP}, the ground coordinate system is denoted by $x_G-y_G$ axes
and the vehicle coordinate system, by $x_V-y_V$ axes.
%
%Let us denote the radar measurement interval or the frame time by $T_f$.
Given angle and range estimates to the scatterers, we first determine the positions of the
scatterers (in the vehicle coordinate system).
Let us define two point sets,
$S(i) = \{{\bf{x}}_k|~ 1\le k \le K \}$ and
$S(i+1) = \{{\bf{z}}_l|~ 1\le l \le L \}$,
where
${\bf{x}}_k \in \mathbb{R}^{2 \times 1}$ and ${\bf{z}}_l \in \mathbb{R}^{2 \times 1}$ denote, respectively
the positions of the scatterers in vehicle coordinates at the measurement of frame times $i$ and $i+1$.
Note that the number of scatterers may vary at every frame.

Let $\Delta {\bf{X}}_G = {\bf X}_G (i+1) - {\bf X}_G (i)$ and
$\Delta \psi = \psi(i+1) - \psi(i)$ be, respectively the changes of the position and angle of the vehicle
in the ground coordinate system, between frames $i$ and $i+1$.
This vehicle motion will induce the corresponding changes in the position vector and the angle
of the scatterer in the vehicle coordinate system,
$\Delta {\bf X}_V$ and $\Delta \theta$.

If we have established the association between the scatterers
${\bf x}_k \in S(i)$ and ${\bf z}_k \in S(i+1)$, $1 \le k \le K$,
then we can find $\Delta \theta$ and $\Delta {\bf X}_V$ by
{\small{
\begin{align}
    (\Delta \theta, \Delta{\bf X}_V) = \mathop {\arg \min }\limits_{\Delta\theta, \Delta{\bf X}_V}
    \sum\limits_{k = 1}^K {||({\bf z}_k - \Delta {\bf X}_V) - {\bf{R}}(\Delta \theta ) {{\bf x}_k} ||^2},
    \label{eq_rigid}
\end{align}
}}
where
\begin{align}
    {\bf{R}}(\Delta \theta ) =
    \left[ {\begin{array}{*{20}{c}}
      {\cos \Delta \theta }&{ - \sin \Delta \theta }\\
       {\sin \Delta \theta }&{\cos \Delta \theta } .
    \end{array}} \right].
\end{align}
On the other hand, If we know $\Delta \theta$ and $\Delta {\bf X}_V$, then we can find the association between
${\bf x}_k$ and ${\bf z}_k$ by
\begin{align}
{{\bf{z}}_k} = \mathop {\arg \min }\limits_{{\bf z}_l \in {S(i + 1)}} ||({\bf{z}}_l - \Delta {\bf X}_V)
- {\bf{R}}(\Delta \theta ){{\bf{x}}_k}|{|^2}.
\label{eq_pointmatching}
\end{align}

However, if $\Delta {\bf X}_V$, $\Delta \theta$ and the association are
all unknown, the above two equations (\ref{eq_rigid}) and (\ref{eq_pointmatching})
need to be iterated until they converge.
%It is well known that this ICP algorithm always converges
This is one of the simplest forms of the ICP method, which is proven to converge
to a local minimum~\cite{Besl1992}.
With the converged $\Delta {\bf X}_V$ and $\Delta \theta$,
the ground coordinate $\Delta {\bf X}_G$ and $\Delta \psi$
can be found by
\begin{align}
\Delta \psi = - \Delta \theta,
\hspace{0.16cm}
\Delta {{\bf{X}}_G} = {{\bf{R}}}(\psi (i + 1)-\pi/2) \cdot (-\Delta {\bf X}_V),
\label{eq_XG}
\end{align}
and the position of the vehicle can be updated as
%\begin{align}
${\bf X}_G (i+1) = {\bf X}_G (i) + \Delta {\bf X}_G$.
%
%Now the incremental vehicle position $\Delta {\bf{X}}_G$ and the angle $\Delta \psi$,
%from frame $i$ to $i+1$
%can be found by using the iterative closest algorithm (ICP) \cite{Besl1992} as
%explained below.
%
%which iterate the (\ref{eq_rigid}) and (\ref{eq_pointmatching}) until they ${\bf{X}}_G(i)$ and $\psi (i)$ converge.
%
%As shown in Fig. \ref{fig_ICP}, the vehicle position and orientation, at $i$-th frame, are denoted
%as ${\bf{X}}_G(i)$ and $\psi (i)$, respectively.
%The vehicle motion can be broken down into two effects: rotation and forward movement.
%Let us assume that the vehicle movement consists of two stages that the vehicle rotates first and subsequently moves straight.
%
%After the vehicle rotates by $\Delta \psi$ and moves forward by $\Delta {\bf{X}}_G$ during frame interval, $T_{f}$,
%the orientation and position at $(i+1)$-th frame become $\psi (i + 1) = \psi (i) + \Delta \psi$
%and ${\bf{X}}_G (i + 1) = {\bf{X}}_G(i) + \Delta {\bf{X}}_G$.
% Fig of Position Update through ICP
%
%%%%%%%%%%%%%%%%%
\subsection{Estimation error in the vehicle localization}
\label{subsec:TrajectoryComparison}

%The scenario of an automatic vehicle parking between two parked cars is depicted in Fig. \ref{sim_PC} (\subref{sim_PointCloud}).

The vehicle motion is modeled as a closed-loop second-order system including a damping effect
and proportional-integral-derivative (PID) controller,
where the vehicle mass and the damping coefficient are assumed to be $1000$ kg and 50 N$\cdot$s/m, respectively.
The control errors in the velocity and orientation angle are zero-mean Gaussian with
$N(0,0.1$ m/s$)$ and $N(0,0.1^o)$, respectively.
The desired reference velocity is $7$ km/h ($=1.95$ m/s) and
the frame interval, $T_{f}$, is $0.01$ s.
%The total number of frames is approximately $400$ in this simulation.

\begin{figure}[!h]
\centering
% sub 1
%\begin{subfigure}{.5\textwidth}
%  \centering
  %\includegraphics[width=\columnwidth]{sim_Trajectory2}
  \includegraphics[width=\columnwidth]{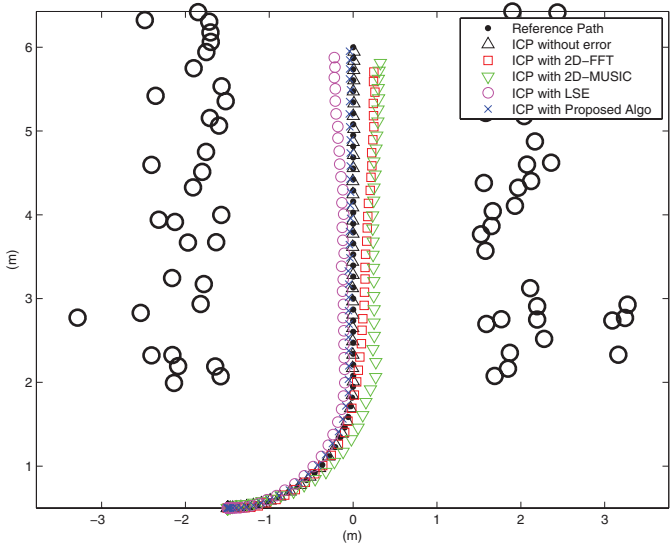}
%  \caption{}
%  \label{sim_Trajectorya}
%\end{subfigure}%
%\hfil
% sub 2
%\begin{subfigure}{.5\textwidth}
%  \centering
%  \includegraphics[width=\columnwidth]{sim_TrajectoryMag}
%  \caption{}
%  \label{sim_TrajectoryMag}
%\end{subfigure}
%
\caption{
The reference path and the actual paths taken by different algorithms.
%(b) The enlarged view of the rectangular area in (a).
}
\label{sim_Trajectory}
\end{figure}

%An automatic vehicle parking was performed according to each estimation method
%, 2D-FFT, 2D-MUSIC, LSE, and the proposed method,
%and we compared the performance of how accurately the vehicle follows along the planned reference path.
%For each scene, the position of the point cloud is assumed to change by $ x \sim N(0,0.01$ m$)$ and $ y \sim N(0,0.01$ m$)$.
%In the ICP algorithm, Welsch criterion function is used.
%
%In the below,
In Fig. \ref{sim_Trajectory},
we compare the proposed estimation algorithm with the others in the autonomous valet parking scenario
using the ICP method under the Welsch criterion function.
The planned reference path is represented as connected-dots.
If the ICP method is applied using error-free positions of the scatterers, then the actual path follows perfectly to the
reference path, which implies that the ICP method itself doesn't incur any error.
With the 2D-FFT or 2D-MUSIC, the bias error accumulates and the actual path deviates from the
reference path by $0.4$ m at the final position, while with the LSE, the final error is about $0.2$ m,
which is still unacceptable for a narrow parking slot.
On the other hand, the proposed algorithm gives only $0.05$ m error at the final position.
%
%
%If the location estimation of the scatterers is performed without error in each scene,
%the actual path entirely depends on the ICP algorithm.
%We confirmed that the actual path follows the reference path perfectly,
%and ICP algorithm itself functions well for this simulation.
%In the estimation with 2D-FFT and 2D-MUSIC, the bias errors are accumulated during
%parking and the actual path does not completely follow the reference path.
%The error is approximately $0.4$ m at the final destination in both cases.
%In the estimation with LSE, even though the small error occur relative to 2D-FFT
%and 2D-MUSIC in each scene, the error at the final destination is about $0.2$ m.
%With proposed method, the actual path is almost same as the reference path
%with $0.05$ m error at the final position.
%
During the parking simulation,
%autonomous parking, t
the total number of frames taken by the 2D-FFT, 2D-MUSIC, LSE, and proposed algorithms
are 389, 401, 385, and 395, respectively.

% Fig of Trajectory
%%%%%%%%%%%%%%%%%
%%%%%%%%%%%%%%%%%

% Table Explanation

%TABLE. \ref{table:RMSEPCTrajectory} shows the information during automatic parking
%%and RMSE of 2D position
%in each estimation method.
%The total number of frames for each estimation method is approximately 400, and the total number of scatterers is 67.
%As the vehicle moves, the number of point cloud inside the FOV may be less than 67.
%The number of detected point cloud may be more than 67 because of an overlapped region of FOV.
%The RMSE of 2D position is calculated for all point cloud obtained for all scenes.
%The RMSE was the largest at $0.0772 (m)$ in the 2D-FFT method and smallest at $0.0280 (m)$ in the proposed method.

%% Table
%\begin{table*}[h]
%\caption{
%During automatic vehicle parking, the comparison of the number of scenes and point cloud
%%, and RMSE over all point clouds and all scenes.
%}
%\label{table:RMSEPCTrajectory}
%\centering
%\begin{tabular}{|l|c|c|c|c|}
%\hline
% & 2D-FFT & 2D-MUSIC & LSE & Proposed \\ \hline
%\# of total frames                & 389      &401      & 385 & 395
%\\ \hline
%\# of scatterers inside FOV (avg,min,max) & 57,36,67 & 57,35,67 & 56,34,67 & 57,36,67 \\ \hline
%\# of scatterers detected (avg,min,max) & 57,36,75 & 57,34,75  & 56,34,74 & 56,36,74
%\\ \hline
%%RMSE of 2D position (m) & 0.0772 & 0.0758 & 0.0330 & 0.0280 \\ \hline
%\end{tabular}
%\end{table*}

%%%%%%%%%%%%%%%%%%%%%%%%%%%%%%%%%%%%%%%%%%%%%%%%%%%%%%%
% Section 6. Conclusion
%%%%%%%%%%%%%%%%%%%%%%%%%%%%%%%%%%%%%%%%%%%%%%%%%%%%%%%
\section{Conclusions}
\label{sec:conclusions}
We have derived the bias error formulas in the range and angle estimates when the conventional
2D-FFT algorithm is applied on the range-angle matrix of the deramped
FMCW radar signals.
We have also proposed a maximum likelihood estimation algorithm that overcomes such bias error
and achieves the CRB.
The proposed algorithm can be used on the range-Doppler matrix or on the angle-Doppler matrix as well.
We have integrated the proposed algorithm with a SLAM problem
to demonstrate its applicability to high precision short-distance autonomous navigation.
%first ML type

%We proposed a novel algorithm for range and angle estimation to eliminate bias error and interference error:
%we considered different time delay for each antenna element depending on angle, and proposed algorithm for multiple targets.
%The proposed method yields improved estimation performance of range and angle, and ultimately attain Cramer-Rao bounds,
%in comparison with conventional approaches, two dimensional Fourier transform (2D-FT) and least square estimation (LSE).

%%%%%%%%%%%%%%%%%%%%%%%%%%%%%%%%%%%%%%%%%%%%%%%%%%%%%%%
% Appendix
%%%%%%%%%%%%%%%%%%%%%%%%%%%%%%%%%%%%%%%%%%%%%%%%%%%%%%%
\appendices
%\appendix {MLE}
%\subsection{Maximum likelihood Estimation}
\section{MLE derivations}
%\subsection{For parameter $\psi_k$}
\label{appendix_ML}
\subsection*{A.1 Parameter $\psi_k$}
Notice that
{\small{
\begin{align}
 \frac{{\partial \Lambda }}{{\partial {\psi _k}}} &=
%\\
- \sum\limits_{m = 0}^{M - 1} {a_k} \left[j{e^{j({\psi _k} + 2\pi \frac{{{u_k}}}{\lambda }m)}}
R_{(k)}^{(m)} - j{e^{ - j({\psi _k} + 2\pi \frac{{{u_k}}}{\lambda }m)}}R{{_{(k)}^{(m)}}^*} \right]
\nonumber
\\
& \hspace{0.3cm} + \sum\limits_{m = 0}^{M - 1} {\sum\limits_{l \ne k}^K {} } {a_k}{a_l}
\left[
 j{e^{j({\psi _k}
+ 2\pi \frac{{{u_k}}}{\lambda }m)}}{e^{ - j({\psi _l} + 2\pi \frac{{{u_l}}}{\lambda }m)}}R_{(l,k)}^{(m)}
\right.
\nonumber
\\
& \left. \hspace{2cm} - j{e^{j({\psi _l} + 2\pi \frac{{{u_l}}}{\lambda }m)}}{e^{ - j({\psi _k}
+ 2\pi \frac{{{u_k}}}{\lambda }m)}}R_{(k,l)}^{(m)}
\right]
\nonumber
\\
& =  - \sum\limits_{m = 0}^{M - 1} {a_k} \left[ j{e^{j({\psi _k} + 2\pi \frac{{{u_k}}}{\lambda }m)}}
S_{(k)}^{(m)} - j{e^{ - j({\psi _k} + 2\pi \frac{{{u_k}}}{\lambda }m)}}S{{_{(k)}^{(m)}}^*} \right]
% \\
%%%%
%%%%
%%%%
\end{align}
\begin{align}
& = 2{a_k}\sum\limits_{m = 0}^{M - 1} {{\mathop{\rm Im}\nolimits} [{e^{j{\psi _k}}}
{e^{j2\pi \frac{{{u_k}}}{\lambda }m}}S_{(k)}^{(m)}]},
\hspace{2cm}
\end{align}
}}
where
{\small{
\begin{gather}
    S_{(k)}^{(m)} = R_{(k)}^{(m)} - \sum\limits_{l \ne k}^K {{a_l}{e^{ - j({\psi _l} + 2\pi \frac{{{u_l}}}
    {\lambda }m)}}R_{(l,k)}^{(m)}}.
\end{gather}
}}%
Therefore, from
$\frac{{\partial \Lambda }}{{\partial {\psi _k}}} = 0$,
we get
{\small{
${\hat \psi _k} =  - \angle (\sum\limits_{m = 0}^{M - 1} {{e^{j2\pi \frac{{{u_k}}}{\lambda }m}}S_{(k)}^{(m)}} )$
}}.

%
%\begin{gather}
%    {\hat \psi _k} =  - \angle (\sum\limits_{m = 0}^{M - 1} {{e^{j2\pi \frac{{{u_k}}}{\lambda }m}}S_{(k)}^{(m)}} ),
%    \label{eq_psi_k}
%\end{gather}
%where the symbol $\angle(\cdot)$ denotes the phase of ($\cdot$).
%
%
\subsection*{A.2 Parameter $a_k$}
Notice that
%Second, we determine $\hat a_k$ by solving $\frac{{\partial \Lambda }}{{\partial {a_k}}} = 0$ for all $k$,
%where
%$\frac{{\partial \Lambda }}{{\partial {a_k}}} $ is expressed as
%
{\small{
\begin{align}
\frac{{\partial \Lambda }}{{\partial {a_k}}} &=
 - \sum\limits_{m = 0}^{M - 1} \left[{e^{j({\psi _k} + 2\pi \frac{{{u_k}}}{\lambda }m)}}R_{(k)}^{(m)} +
{e^{ - j({\psi _k} + 2\pi \frac{{{u_k}}}{\lambda }m)}}R{{_{(k)}^{(m)}}^*} \right]
\nonumber
\\
& \hspace{0.3cm} + \sum\limits_{m = 0}^{M - 1} {\sum\limits_{l \ne k}^K {} } {a_l} \left[{e^{j({\psi _k}
+ 2\pi \frac{{{u_k}}}{\lambda }m)}}{e^{ - j({\psi _l} + 2\pi \frac{{{u_l}}}{\lambda }m)}}R_{(l,k)}^{(m)}
\right.
\nonumber
\\
&
\hspace{2cm}
\left.
+ {e^{j({\psi _l} + 2\pi \frac{{{u_l}}}{\lambda }m)}}{e^{ - j({\psi _k}
+ 2\pi \frac{{{u_k}}}{\lambda }m)}}R_{(k,l)}^{(m)}\right]
\nonumber
\\
& \hspace{0.3cm} + 2MN{a_k}
%\nonumber
\\
%\\
& =  - 2\sum\limits_{m = 0}^{M - 1} {{\mathop{\rm Re}\nolimits} \left[{e^{j({\psi _k} + 2\pi \frac{{{u_k}}}{\lambda }m)}}
R_{(k)}^{(m)}\right]}
\nonumber
\\
& \hspace{0.3cm}
+ 2\sum\limits_{m = 0}^{M - 1} {\sum\limits_{l \ne k}^K {{a_l}{\mathop{\rm Re}\nolimits}
\left[{e^{j({\psi _k} + 2\pi \frac{{{u_k}}}{\lambda }m)}}{e^{ - j({\psi _l}
+ 2\pi \frac{{{u_l}}}{\lambda }m)}}R_{(l,k)}^{(m)}\right]} }
\nonumber
\\
& \hspace{0.3cm} + 2MN{a_k}.
\end{align}
}}
Therefore, from
%Then, to satisfy
$\frac{{\partial \Lambda }}{{\partial {a_k}}} = 0$,
%the $\hat a_k$ should be
we get
{\small{
\begin{align}
& {\hat a_k} = \frac{1}{{MN}} \left[\sum\limits_{m = 0}^{M - 1} {{\mathop{\rm Re}\nolimits}
[{e^{j({\psi _k} + 2\pi \frac{{{u_k}}}{\lambda }m)}}R_{(k)}^{(m)}]}
\right.
\nonumber
\\
& \hspace{0.3cm}
\left.
- \sum\limits_{m = 0}^{M - 1} {\sum\limits_{l \ne k}^K {{a_l}{\mathop{\rm Re}\nolimits}
[{e^{j({\psi _k} + 2\pi \frac{{{u_k}}}{\lambda }m)}}{e^{ - j({\psi _l}
+ 2\pi \frac{{{u_l}}}{\lambda }m)}}R_{(l,k)}^{(m)}]} } \right].
\end{align}
}}
The matrix form of this equation is (\ref{eq_a_k}).
%
%\begin{align}
%    {\bf{B}} \cdot {\bf{a}} = {\bf{y}},
%    \label{eq_a_k}
%\end{align}
%where
%${\bf{a}} = {[{\hat a_1},...,{\hat a_K}]^T} \in \mathbb{R}^{K \times 1} $ ,
%${\bf{B}} \in \mathbb{R}^{K \times K} $ ,
%and ${\bf{y}} \in \mathbb{R}^{K \times 1} $.
%The components of ${\bf{B}}$ and ${\bf{y}}$ are
%\begin{align}
%&{\bf{B}}(k,k) = MN,
%\\
%&{\bf{B}}(k,l) = \sum\limits_{m = 0}^{M - 1} {{\mathop{\rm Re}\nolimits} [{e^{j({\psi _k} + 2\pi \frac{{{u_k}}}{\lambda }m)}}{e^{ - j({\psi _l} + 2\pi \frac{{{u_l}}}{\lambda }m)}}R_{(l,k)}^{(m)}]},
%\\
%&{\bf{y}}(k) = \sum\limits_{m = 0}^{M - 1} {{\mathop{\rm Re}\nolimits} [{e^{j({\psi _k} + 2\pi \frac{{{u_k}}}{\lambda }m)}}R_{(k)}^{(m)}]},
%\end{align}
%where
%$k=1,...,K$, $l=1,...,K$, and $l \ne k$.

\subsection*{A.3 Parameter $u_k$}

%Third, we solve $\frac{{\partial \Lambda }}{{\partial {u_k}}} = 0$ to determine $\hat u_k$.
%We solve a zero-finding problem with one-step of Newton's method,
Jacobian of $\Lambda(\cdot)$ with respect to $u_k$ is as follows.
%
%\begin{gather}
%    {\hat u_k}(i + 1) = {\hat u_k}(i) - \frac{f_u}{{f_u'}}{|_{{\hat u_k} = {\hat u_k}(i)}},
%\end{gather}
%where $\hat u_k(i)$ denotes $\hat u_k$ at $i$-th iteration.
%We can derive $f_u$ and $f_u'$ as follows.
{\small{
\begin{align}
f_u &= \frac{{\partial \Lambda }}{{\partial {u_k}}} =
%\\
%&
- {a_k}\sum\limits_{m = 0}^{M - 1} {} \left[ j\frac{{2\pi }}{\lambda }m{e^{j({\psi _k}
+ 2\pi \frac{{{u_k}}}{\lambda }m)}}R_{(k)}^{(m)}
\right.
\nonumber
\\
& \hspace{1cm} - j\frac{{2\pi }}
{\lambda }m{e^{ - j({\psi _k} + 2\pi \frac{{{u_k}}}{\lambda }m)}}R_{(k)}^{(m)*}
\nonumber
\\
& \hspace{1cm} + \left.{e^{j({\psi _k} + 2\pi \frac{{{u_k}}}{\lambda }m)}}
\frac{{\partial R_{(k)}^{(m)}}}{{\partial {u_k}}} + {e^{ - j({\psi _k} + 2\pi
\frac{{{u_k}}}{\lambda }m)}}\frac{{\partial R_{(k)}^{(m)*}}}{{\partial {u_k}}}\right]
\nonumber
\\
& \hspace{0cm} + {a_k}\sum\limits_{m = 0}^{M - 1} {\sum\limits_{l \ne k}^K {} } \left[{a_l}(j\frac{{2\pi }}{\lambda })m{e^{j({\psi _k}
+ 2\pi \frac{{{u_k}}}{\lambda }m)}}{e^{ - j({\psi _l} + 2\pi \frac{{{u_l}}}{\lambda }m)}}R_{(l,k)}^{(m)} \right.
\nonumber
\\
& \hspace{1cm} + {a_l}{e^{j({\psi _k} + 2\pi \frac{{{u_k}}}{\lambda }m)}}{e^{ - j({\psi _l}
+ 2\pi \frac{{{u_l}}}{\lambda }m)}}\frac{{\partial R_{(l,k)}^{(m)}}}{{\partial {u_k}}}
\nonumber
\\
& \hspace{1cm} + {a_l}( - j\frac{{2\pi }}{\lambda })m{e^{ - j({\psi _k} + 2\pi \frac{{{u_k}}}{\lambda }m)}}
{e^{j({\psi _l} + 2\pi \frac{{{u_l}}}{\lambda }m)}}R_{(l,k)}^{(m)*}
\nonumber
\\
& \hspace{1cm} + \left.{a_l}{e^{ - j({\psi _k} + 2\pi \frac{{{u_k}}}{\lambda }m)}}
{e^{j({\psi _l} + 2\pi \frac{{{u_l}}}{\lambda }m)}}\frac{{\partial R_{(l,k)}^{(m)*}}}{{\partial {u_k}}} \right]
%\\
\end{align}
\begin{align}
& = \frac{{4\pi }}{\lambda }{a_k}\sum\limits_{m = 0}^{M - 1} {} m  {\mathop{\rm Im}\nolimits}
[{e^{j({\psi _k} + 2\pi \frac{{{u_k}}}{\lambda }m)}}R_{(k)}^{(m)}]
\nonumber
\\
& \hspace{0.3cm} - 2{a_k}\sum\limits_{m = 0}^{M - 1} {} {\mathop{\rm Re}\nolimits} [{e^{j({\psi _k} + 2\pi
\frac{{{u_k}}}{\lambda }m)}}\frac{{\partial R_{(k)}^{(m)}}}{{\partial {u_k}}}]
\nonumber
\\
& \hspace{0.3cm} - \frac{{4\pi }}{\lambda }{a_k}\sum\limits_{m = 0}^{M - 1} {\sum\limits_{l \ne k}^K {} } m
{\mathop{\rm Im}\nolimits} [{e^{j({\psi _k} + 2\pi \frac{{{u_k}}}{\lambda }m)}}{a_l}{e^{ - j({\psi _l}
+ 2\pi \frac{{{u_l}}}{\lambda }m)}}R_{(l,k)}^{(m)}]
\nonumber
\\
& \hspace{0.3cm} + 2{a_k}\sum\limits_{m = 0}^{M - 1} {\sum\limits_{l \ne k}^K {{a_l}{\mathop{\rm Re}\nolimits}
[{e^{j({\psi _k} + 2\pi \frac{{{u_k}}}{\lambda }m)}}{e^{ - j({\psi _l} + 2\pi \frac{{{u_l}}}{\lambda }m)}}
\frac{{\partial R_{(l,k)}^{(m)}}}{{\partial {u_k}}}]} }
%\nonumber
\\
& = \frac{{4\pi }}{\lambda }{a_k}\sum\limits_{m = 0}^{M - 1} {} m  {\mathop{\rm Im}\nolimits}
[{e^{j({\psi _k} + 2\pi \frac{{{u_k}}}{\lambda }m)}}S_{(k)}^{(m)}]
\nonumber
\\
& \hspace{0.3cm} - 2{a_k}\sum\limits_{m = 0}^{M - 1} {} {\mathop{\rm Re}\nolimits} [{e^{j({\psi _k}
+ 2\pi \frac{{{u_k}}}{\lambda }m)}}\frac{{\partial S_{(k)}^{(m)}}}{{\partial {u_k}}}],
\end{align}
}}%
where
{\small{
\begin{align}
    %S_{(k)}^{(m)} & = R_{(k)}^{(m)} - \sum\limits_{l \ne k}^K {{a_l}{e^{ - j({\psi _l}
    %+ 2\pi \frac{{{u_l}}}{\lambda }m)}}R_{(l,k)}^{(m)}},
    %\\
    \frac{{\partial S_{(k)}^{(m)}}}{{\partial {u_k}}} & = \frac{{\partial R_{(k)}^{(m)}}}{{\partial {u_k}}}
    - \sum\limits_{l \ne k}^K {{a_l}{e^{ - j({\psi _l} + 2\pi \frac{{{u_l}}}{\lambda }m)}}
    \frac{{\partial R_{(l,k)}^{(m)}}}{{\partial {u_k}}}},
    \label{eq_S_u}
    \\
    \frac{{\partial R_{(k)}^{(m)}}}{{\partial {u_k}}} & = \sum\limits_{n = 0}^{N - 1}
    {{z^*}[n,m] \left[j\frac{{2\pi B}}{{cN}}mn\right] {e^{j2\pi (2{r_k} + m{u_k})\frac{B}{{cN}}n}}},
    \\
    \frac{{\partial R_{(l,k)}^{(m)}}}{{\partial {u_k}}} & = \sum\limits_{n = 0}^{N - 1}
    {\left[j\frac{{2\pi B}}{{cN}}mn\right]{e^{j2\pi (2({r_k} - {r_l}) + m({u_k} - {u_l}))\frac{B}{{cN}}n}}}.
\end{align}
}}
The Hessian of $\Lambda(\cdot)$ with respect to $u_k$ is given by
{\small{
\begin{align}
    f_u' & = %\frac{{{\partial ^2}\Lambda }}{{\partial {u_k}^2}} =
    \frac{\partial }{{\partial {u_k}}}
    (\frac{{\partial \Lambda }}{{\partial {u_k}}})
    = \frac{{4\pi }}{\lambda }{a_k}\sum\limits_{m = 0}^{M - 1} {} m\frac{\partial }{{\partial {u_k}}}
    {\mathop{\rm Im}\nolimits} [{e^{j({\psi _k} + 2\pi \frac{{{u_k}}}{\lambda }m)}}S_{(k)}^{(m)}]
\nonumber
    \\
    & \hspace{0.3cm} - 2{a_k}\sum\limits_{m = 0}^{M - 1} {} \frac{\partial }{{\partial {u_k}}}
    {\mathop{\rm Re}\nolimits} [{e^{j({\psi _k} + 2\pi \frac{{{u_k}}}{\lambda }m)}}\frac{{\partial S_{(k)}^{(m)}}}
    {{\partial {u_k}}}]
%\nonumber
    \\
    & = 2{(\frac{{2\pi }}{\lambda })^2}{a_k}\sum\limits_{m = 0}^{M - 1} {} {m^2}{\mathop{\rm Re}\nolimits}
    [{e^{j({\psi _k} + 2\pi \frac{{{u_k}}}{\lambda }m)}}S_{(k)}^{(m)}]
\nonumber
    \\
    & \hspace{0.3cm} + 4(\frac{{2\pi }}{\lambda }){a_k}\sum\limits_{m = 0}^{M - 1} {} m{\mathop{\rm Im}\nolimits}
    [{e^{j({\psi _k} + 2\pi \frac{{{u_k}}}{\lambda }m)}}\frac{{\partial S_{(k)}^{(m)}}}{{\partial {u_k}}}]
    \nonumber
    \\
    & \hspace{0.3cm}
    - 2{a_k}\sum\limits_{m = 0}^{M - 1} {} {\mathop{\rm Re}\nolimits} \left[{e^{j({\psi _k}
    + 2\pi \frac{{{u_k}}}{\lambda }m)}}\frac{{{\partial ^2}S_{(k)}^{(m)}}}{{\partial {u_k}^2}}\right],
\end{align}
}}%
where
{\small{
\begin{align}
    & \frac{{{\partial ^2}S_{(k)}^{(m)}}}{{\partial {u_k}^2}} = \frac{{{\partial ^2}R_{(k)}^{(m)}}}
    {{\partial {u_k}^2}} - \sum\limits_{l \ne k}^K {{a_l}{e^{ - j({\psi _l} + 2\pi \frac{{{u_l}}}
    {\lambda }m)}}\frac{{{\partial ^2}R_{(l,k)}^{(m)}}}{{\partial {u_k}^2}}},
    \label{eq_S_u2}
    \\
    & \frac{{{\partial ^2}R_{(k)}^{(m)}}}{{\partial {u_k}^2}} =  - \sum\limits_{n = 0}^{N - 1}
    {{z^*}[n,m]{{(\frac{{2\pi B}}{{cN}}mn)}^2}{e^{j2\pi (2{r_k} + m{u_k})\frac{B}{{cN}}n}}},
    \\
    & \frac{{{\partial ^2}R_{(l,k)}^{(m)}}}{{\partial {u_k}^2}} =  - \sum\limits_{n = 0}^{N - 1}
    {{{(\frac{{2\pi B}}{{cN}}mn)}^2}{e^{j2\pi (2({r_k} - {r_l}) + m({u_k} - {u_l}))\frac{B}{{cN}}n}}}.
\end{align}
}}
We note that
{\small{
\begin{align}
  & \frac{\partial }{{\partial {u_k}}}{\mathop{\rm Im}\nolimits} [{e^{j({\psi _k}
  + 2\pi \frac{{{u_k}}}{\lambda }m)}}S_{(k)}^{(m)}]
  =
  \frac{{2\pi }}{\lambda }m{\mathop{\rm Re}\nolimits} [{e^{j({\psi _k}
  + 2\pi \frac{{{u_k}}}{\lambda }m)}}S_{(k)}^{(m)}]
  \nonumber
  \\
  & \hspace{3.5cm}
  + {\mathop{\rm Im}\nolimits} [{e^{j({\psi _k}
  + 2\pi \frac{{{u_k}}}{\lambda }m)}}\frac{{\partial S_{(k)}^{(m)}}}{{\partial {u_k}}}],
  \end{align}
  \begin{align}
%  \\
  & \frac{\partial }{{\partial {u_k}}}{\mathop{\rm Re}\nolimits} [{e^{j({\psi _k} + 2\pi \frac{{{u_k}}}
  {\lambda }m)}}\frac{{\partial S_{(k)}^{(m)}}}{{\partial {u_k}}}] =
  \nonumber
  \\
  & \hspace{0cm}
  - \frac{{2\pi }}{\lambda }m{\mathop{\rm Im}\nolimits} [{e^{j({\psi _k}
  + 2\pi \frac{{{u_k}}}{\lambda }m)}}\frac{{\partial S_{(k)}^{(m)}}}{{\partial {u_k}}}]
  + {\mathop{\rm Re}\nolimits} [{e^{j({\psi _k} + 2\pi \frac{{{u_k}}}{\lambda }m)}}\frac{{{\partial ^2}
  S_{(k)}^{(m)}}}{{\partial {u_k}^2}}].
\end{align}
}}

\subsection*{A.4 Parameter $r_k$}
Jacobian of $\Lambda(\cdot)$ with respect to $r_k$ is as follows.
{\small{
\begin{align}
& f_r = \frac{{\partial \Lambda }}{{\partial {r_k}}} =
%&
- \sum\limits_{m = 0}^{M - 1} {} \left[{a_k}{e^{j({\psi _k} + 2\pi \frac{{{u_k}}}{\lambda }m)}}
\frac{{\partial R_{(k)}^{(m)}}}{{\partial {r_k}}}
\right.
\nonumber
\\
& \hspace{2cm}
\left.
+ {a_k}{e^{ - j({\psi _k} + 2\pi \frac{{{u_k}}}{\lambda }m)}}
\frac{{\partial R_{(k)}^{(m)*}}}{{\partial {r_k}}} \right]
\nonumber
\\
& \hspace{0.3cm} + \sum\limits_{m = 0}^{M - 1} {\sum\limits_{l \ne k}^K {} } \left[{a_k}{a_l}{e^{j({\psi _k}
+ 2\pi \frac{{{u_k}}}{\lambda }m)}}{e^{ - j({\psi _l} + 2\pi \frac{{{u_l}}}{\lambda }m)}}
\frac{{\partial R_{(l,k)}^{(m)}}}{{\partial {r_k}}}
\right.
\nonumber
\\
& \hspace{1.3cm}
\left.
+ {a_k}{a_l}{e^{ - j({\psi _k} + 2\pi \frac{{{u_k}}}{\lambda }m)}}{e^{j({\psi _l} + 2\pi \frac{{{u_l}}}
{\lambda }m)}}\frac{{\partial R_{(l,k)}^{(m)*}}}{{\partial {r_k}}}\right]
%\nonumber
\\
& =  - 2{a_k}\sum\limits_{m = 0}^{M - 1} {} {\mathop{\rm Re}\nolimits} [{e^{j({\psi _k}
+ 2\pi \frac{{{u_k}}}{\lambda }m)}}\frac{{\partial S_{(k)}^{(m)}}}{{\partial {r_k}}}],
\end{align}
}}%
where
{\small{
\begin{align}
    \frac{{\partial S_{(k)}^{(m)}}}{{\partial {r_k}}} & = \frac{{\partial R_{(k)}^{(m)}}}
    {{\partial {r_k}}} - \sum\limits_{l \ne k}^K {{a_l}{e^{ - j({\psi _l} + 2\pi
    \frac{{{u_l}}}{\lambda }m)}}\frac{{\partial R_{(l,k)}^{(m)}}}{{\partial {r_k}}}},
    \label{eq_S_r}
    \\
    \frac{{\partial R_{(k)}^{(m)}}}{{\partial {r_k}}} & = \sum\limits_{n = 0}^{N - 1}
    {{z^*}[n,m]\left[j\frac{{4\pi B}}{{cN}}n\right]{e^{j2\pi (2{r_k} + m{u_k})\frac{B}{{cN}}n}}},
    \\
    \frac{{\partial R_{(l,k)}^{(m)}}}{{\partial {r_k}}} & = \sum\limits_{n = 0}^{N - 1}
    {\left[j\frac{{4\pi B}}{{cN}}n\right]{e^{j2\pi (2({r_k} - {r_l}) + m({u_k} - {u_l}))\frac{B}{{cN}}n}}} .
\end{align}
}}
The Hessian of $\Lambda(\cdot)$ with respect to $r_k$ is given by
{\small{
\begin{align}
f_r' & %= \frac{{{\partial ^2}\Lambda }}{{\partial {r_k}^2}} =
=\frac{\partial }{{\partial {r_k}}} (\frac{{\partial \Lambda }}{{\partial {r_k}}})
%\\
%&
= - 2{a_k}\sum\limits_{m = 0}^{M - 1} {} {\mathop{\rm Re}\nolimits}
\left[{e^{j({\psi _k} + 2\pi \frac{{{u_k}}}{\lambda }m)}}\frac{{{\partial ^2}S_{(k)}^{(m)}}}{{\partial {r_k}^2}}\right],
\end{align}
}}%
where
{\small{
\begin{align}
    & \frac{{{\partial ^2}S_{(k)}^{(m)}}}{{\partial {r_k}^2}} = \frac{{{\partial ^2}R_{(k)}^{(m)}}}
    {{\partial {r_k}^2}} - \sum\limits_{l \ne k}^K {{a_l}{e^{ - j({\psi _l} + 2\pi \frac{{{u_l}}}{\lambda }m)}}
    \frac{{{\partial ^2}R_{(l,k)}^{(m)}}}{{\partial {r_k}^2}}},
%    \label{eq_S_r2}
    \\
    & \frac{{{\partial ^2}R_{(k)}^{(m)}}}{{\partial {r_k}^2}} =  - \sum\limits_{n = 0}^{N - 1}
    {{z^*}[n,m]{{\left[\frac{{4\pi B}}{{cN}}n\right]}^2}{e^{j2\pi (2{r_k} + m{u_k})\frac{B}{{cN}}n}}},
    \\
    & \frac{{{\partial ^2}R_{(l,k)}^{(m)}}}{{\partial {r_k}^2}} =  - \sum\limits_{n = 0}^{N - 1}
    {{{\left[\frac{{4\pi B}}{{cN}}n\right]}^2}{e^{j2\pi (2({r_k} - {r_l}) + m({u_k} - {u_l}))\frac{B}{{cN}}n}}}.
\end{align}
}}

\section{Cramer-Rao Bounds}
\label{sec:CRB}
Let
$\text{\boldmath$\omega$}  = [\omega_1, \omega_2, \omega_3, \omega_4]^T = {[a,\,\psi ,\,r,\,u]^T}$. Then the
covariance matrix ${\bf{C}}_{\hat \omega} \in \mathbb{R}^{4 \times 4}$
of an unbiased estimator ${\hat {\bf{\omega}}}$ has the lower bound
%\begin{gather}
${{\bf{C}}_{\hat \omega }} \ge {{\bf{I}}^{ - 1}} (\text{\boldmath$\omega$})$.
%\end{gather}
%
For our measurement model,
%where $\omega  = [\omega_1, \omega_2, \omega_3, \omega_4] =
%{[a,\,\psi ,\,r,\,u]^T} $ and ${\bf{I}}(\omega )$ is $(4 \times 4)$ Fisher information matrix (FIM). We recall that
%the signal model,
%%
%\begin{gather}
%    z[n,m] = s[n,m] + w[n,m],
%\end{gather}
%%
%where
\begin{gather}
    z[n,m] = s[n,m] + w[n,m],
\end{gather}
where
$s[n,m] = a{e^{j(\psi  + 2\pi \frac{u}{\lambda }m + 2\pi (2r + mu)\frac{B}{{cN}}n)}}$,
and ${\bf{I}}(\text{\boldmath$\omega$})$ is given by
%Since $w[n,m]$ is temporarily and spatially uncorrelated complex Gaussian noise, ($i$,$j$)-th component of FIM is as follows.
%
{\small{
\begin{align}
    {[{\bf{I}}(\text{\boldmath$\omega$})]_{i,j}} = \frac{2}{{{\sigma ^2}}}
    \sum\limits_{m = 0}^{M - 1} {\sum\limits_{n = 0}^{N - 1}{}}
    %\label{eq_FIM_component}
    %\\
    {{\left[\frac{{\partial \mu [n,m]}}{{\partial {\omega _i}}}\frac{{\partial \mu [n,m]}}{{\partial {\omega _j}}}
    + \frac{{\partial \nu [n,m]}}{{\partial {\omega _i}}}\frac{{\partial \nu [n,m]}}{{\partial {\omega _j}}}\right]} },
\end{align}
}}%
where $\mu [n,m] = {\mathop{\rm Re}\nolimits} [s[n,m]]$ and $\nu [n,m] = {\mathop{\rm Im}\nolimits} [s[n,m]]$.
Therefore, with the definition,
%\begin{align}
$h[n,m] = \psi  + 2\pi \frac{u}{\lambda }m + 2\pi (2r + mu)\frac{B}{{cN}}n$,
%\end{align}
we get
%
%First, we derive $\frac{{\partial \mu [n,m]}}{{\partial {\omega _i}}}$ and
%$\frac{{\partial \nu [n,m]}}{{\partial {\omega _i}}}$ for $i=1,..,4$.
%
{\small{
\begin{align}
&\frac{{\partial \mu [n,m]}}{{\partial {\omega _1}}}  = \cos (h[n,m]), \quad
\frac{{\partial \mu [n,m]}}{{\partial {\omega _2}}}  =  - a\sin (h[n,m]),
%\nonumber
\\
&\frac{{\partial \mu [n,m]}}{{\partial {\omega _3}}}  =  - 4\pi a\frac{B}{{cN}}n\sin (h[n,m]),
%\nonumber
\\
& \frac{{\partial \mu [n,m]}}{{\partial {\omega _4}}}  =  - 2\pi am(\frac{1}{\lambda } + \frac{B}{{cN}}n)\sin (h[n,m]),
%\nonumber
\\
%\end{align}
%
%\begin{align}
&\frac{{\partial \nu [n,m]}}{{\partial {\omega _1}}}  = \sin (h[n,m]), \quad
\frac{{\partial \nu [n,m]}}{{\partial {\omega _2}}}  = a\cos (h[n,m]),
%\nonumber
\\
&\frac{{\partial \nu [n,m]}}{{\partial {\omega _3}}}  = 4\pi a\frac{B}{{cN}}n\cos (h[n,m]),
%\nonumber
\\
& \frac{{\partial \nu [n,m]}}{{\partial {\omega _4}}}  = 2\pi am(\frac{1}{\lambda } + \frac{B}{{cN}}n)\cos (h[n,m]).
\end{align}
}}%
Therefore,
{\small{
\begin{align}
{\bf{I}}(\text{\boldmath$\omega$}) = \frac{2}{{{\sigma ^2}}}\sum\limits_{m = 0}^{M - 1} {\sum\limits_{n = 0}^{N - 1} {} }
\left[
{\begin{array}{*{20}{c}}
1 & 0 & 0 & 0
\\
0 &
{{a^2}} &
{4\pi {a^2}\frac{B}{{cN}}n} &
i_{2,4}
\\
0 &
{4\pi {a^2}\frac{B}{{cN}}n} &
i_{3,3} &
i_{3,4}
\\
0&
i_{4,2} &
i_{4,3} &
i_{4,4}
\end{array}}
\right],
\label{eq_FIM}
%\nonumber
\end{align}
}}%
where
{\small{
\begin{align}
    & i_{2,4} = i_{4,2} = {2\pi {a^2}m \left[\frac{1}{\lambda } + \frac{B}{{cN}}n \right]},
    \quad
%    \nonumber \\
    i_{3,3} = \left[ 4\pi a\frac{B}{{cN}}n \right]^2,
%    \nonumber
    \\
    & i_{3,4} = i_{4,3} = {2{{(2\pi a)}^2}\left[\frac{1}{\lambda }
    + \frac{B}{{cN}}n\right]\frac{B}{{cN}}mn},
    \\
    & i_{4,4} = {{{(2\pi am)}^2}{\left[\frac{1}{\lambda } + \frac{B}{{cN}}n\right]^2}}.
\end{align}
}}
We can easily find the $3$rd and $4$th diagonal entries of ${{\bf{I}}^{ - 1}}(\text{\boldmath$\omega$})$, using
the Cramer's rule
%
%by putting the parameters in (\ref{eq_FIM}) and taking an inverse matrix.
%Then, we can determine the minimum deviation of
to get the CRB of $r$ and $\theta$;
\begin{align}
& {\sigma _{r,CRB}} = \sqrt {{{[{{\bf{I}}^{ - 1}}(\text{\boldmath$\omega$})]}_{3,3}}}
\\
& {\sigma _{\theta ,CRB}} = \sqrt {{{[{{\bf{I}}^{ - 1}}(\text{\boldmath$\omega$} )]}_{4,4}}} /\frac{{du}}{{d\theta }}
= \sqrt {{{[{{\bf{I}}^{ - 1}}(\text{\boldmath$\omega$} )]}_{4,4}}} /(d\cos \theta ).
\end{align}
where $u = d\sin {\theta}$, and the unit of ${\sigma _{\theta ,CRB}}$ is radian.

%\section*{Acknowledgment}


\ifCLASSOPTIONcaptionsoff
  \newpage
\fi



%\bibliographystyle{IEEEtran}
%\begin{thebibliography}{1}


%\end{thebibliography}

%\bibliographystyle{IEEEtran}
%\bibliography{refs}
%\bibliographystyle{plain}
\bibliographystyle{IEEEtran}
\bibliography{sensors}

%\bibliography{strings,refs}


%\end{IEEEbiography}
%%%%%%%%%%%%%%%%%%%%%%%%%%%%%%%%%%%%%%%%%


%\newpage



%\vfill


\end{document}